\documentclass[11pt]{revtex4-2}

\usepackage{graphicx}% Include figure files
\usepackage{dcolumn}% Align table columns on decimal point
\usepackage{bm}% bold math
\usepackage[table]{xcolor}
\usepackage{graphicx}
\usepackage{adjustbox}
\usepackage{placeins}
\usepackage[T1]{fontenc}
\usepackage{lipsum}
\usepackage{csquotes}
\usepackage{colortbl}
\usepackage{hyperref}
\usepackage{bm}
\usepackage{multirow}
\usepackage{booktabs}
\usepackage{epsfig}
\usepackage{amsmath,amssymb}

\begin{document}

\title{Incorporating Long-Range Interactions via the Multipole Expansion into Ground and Excited-State Molecular Simulations}

\author{Rhyan Barrett}
\affiliation{Leipzig University, Wilhelm Ostwald Institute for Physical
and Theoretical Chemistry, Linnéstraße 2, 04103 Leipzig, Germany}
\author{Johannes C. B. Dietschreit}
\affiliation{Institute of Theoretical Chemistry, Faculty of Chemistry, University of Vienna, Währinger Straße 17, 1090 Vienna, Austria}
\author{Julia Westermayr}
\email{julia.westermayr@uni-leipzig.de}
\affiliation{Leipzig University, Wilhelm Ostwald Institute for Physical
and Theoretical Chemistry, Linnéstraße 2, 04103 Leipzig, Germany}
\affiliation{Center for Scalable Data Analytics and Artificial Intelligence (ScaDS.AI), Dresden/Leipzig, Germany}

\date{\today}% It is always \today, today,

\begin{abstract}
Simulating long-range interactions remains a significant challenge for molecular machine learning potentials due to the need to accurately capture interactions over large spatial regions. In this work, we introduce FieldMACE, an extension of the message-passing atomic cluster expansion (MACE) architecture that integrates the multipole expansion to model long-range interactions  more efficiently. By incorporating the multipole expansion, FieldMACE effectively captures environmental and long-range effects in both ground and excited states. Benchmark evaluations demonstrate its superior performance in predictions and computational efficiency compared to previous architectures, as well as its ability to accurately simulate nonadiabatic excited-state dynamics. Furthermore, transfer learning from foundational models enhances data efficiency, making FieldMACE a scalable, robust, and transferable framework for large-scale molecular simulations.

\end{abstract}

\maketitle
%%%FOOTNOTES%%%

\section{Introduction}

Over the past few years, graph neural networks (GNNs) have shown great promise in modeling molecules and materials due to the natural graph-like structure of chemical systems \cite{sauceda2022bigdml, deng2023chgnet, gao2020torchani, ko2021fourth}. However, capturing long-range interactions, particularly electrostatics, remains a key challenge \cite{anstine2023machine}. These interactions are crucial when considering solvent effects \cite{Reichardt2007,Reichardt2011} or for phenomena such as protein folding \cite{Go1978, Sagui1999}. However, their slow decay with distance makes them difficult to represent accurately \cite{Ambrosetti2014, Kang2024}.

Many frequently applied GNN models focus on interactions among immediate neighbors \cite{batzner20223, schutt2021equivariant}. As a result, purely local descriptors often miss subtle, but important long-range forces, leading to inaccuracies when simulating systems like proteins or hybrid organic-inorganic interfaces \cite{Gromiha1999, westermayr2022long} that are often dominated by these effects. Various strategies have been proposed to address this limitation \cite{kosmala2023ewald, loche2024fast}, among them, message-passing methods, which exchange information among neighboring nodes, but their limited depth restricts the effective receptive field, making it difficult to account for distant atoms. Other approaches use additional physics-based terms \cite{Anstine2023,Gao2022} or incorporate data from large, extended systems \cite{behler2021machine, kosmala2023ewald} such as some of the systems in the OC20 \cite{chanussot2021open}
dataset, which describes molecules on surfaces, although these can be prohibitively expensive to generate.

\begin{figure*}[!htbp]
    \centering
    \includegraphics[width=0.75\textwidth]{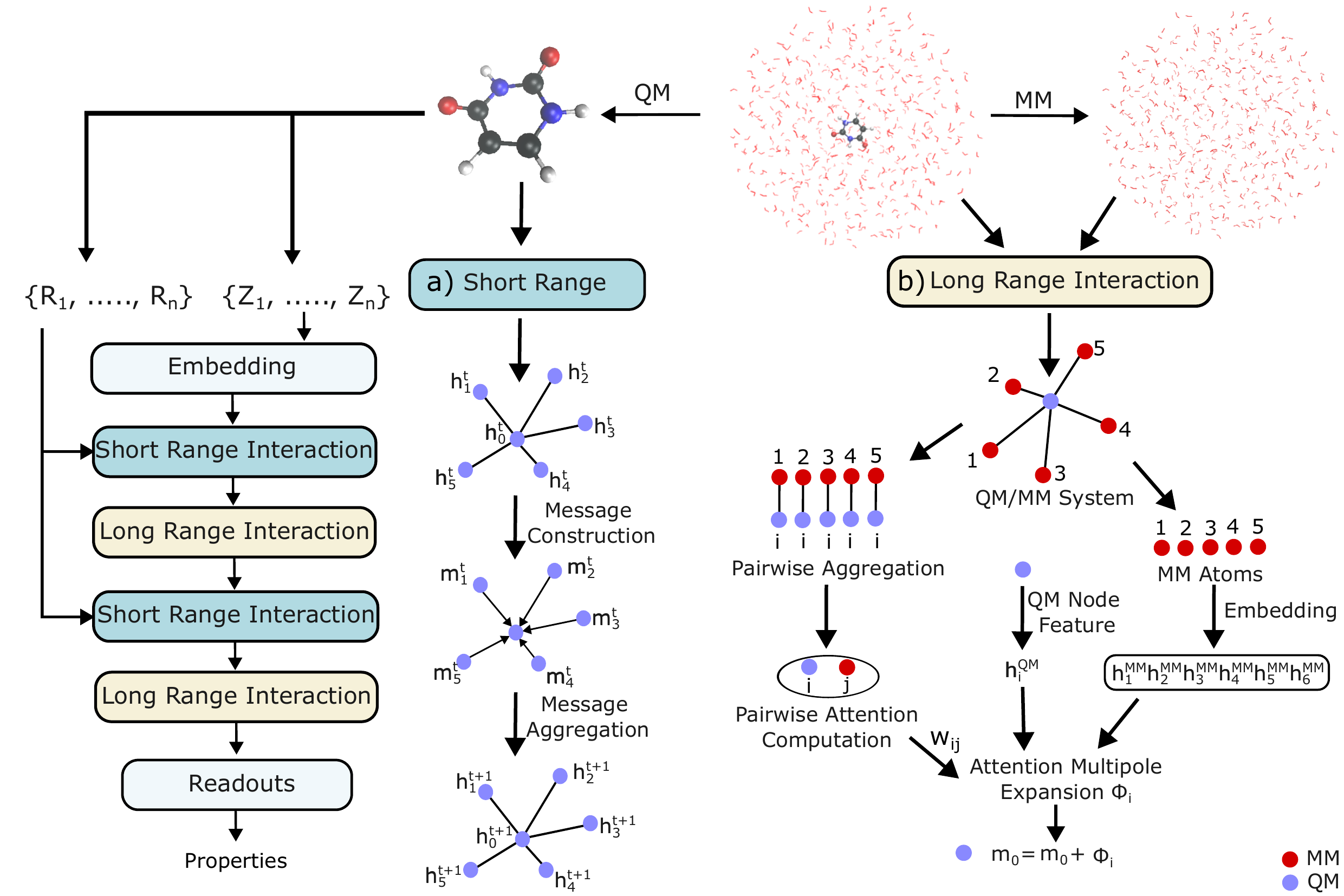}

    \caption{\textbf{Architecture of the message-passing neural network integrated with the multipole expansion} a) The short-range blocks operate through a message-passing scheme aggregating information in the atoms local neighborhood, whereas the long-range blocks collect information on molecular mechanics (MM) nodes through the multipole expansion.  b) 
  The long-range block uses pairwise attention weightings that are computed between the quantum mechanics (QM) region and MM atoms. These are combined with MM node features and the QM node features in the multipole expansion described in Section~\ref{sec:MPE}.
    }
    \label{fig:method}
\end{figure*}

A more efficient alternative takes inspiration from QM/MM (quantum mechanics/molecular mechanics) \cite{tzeliou2022review,boereboom2018explicit} methodologies. In these methods, the region undergoing critical chemical transformations is modeled quantum mechanically, while the surrounding environment is treated classically. An illustration of this approach is shown in the top right of Fig.~\ref{fig:method}. This reduces the computational cost for large systems without sacrificing essential chemical accuracy.

Traditional machine learning (ML) models that include long-range effects often assign node features to every atom in the system and perform periodic updates thereof, even if many atoms are only characterized by a simple charge \cite{frank2022so3krates, batatia2023equivariant, Frank2024}. This leads to excessive computational overhead and poor scaling. To address this issue, some approaches incorporate physics-based features from the classically treated region into the quantum region. For example, FieldSchNet \cite{gastegger2021machine} includes the electric field generated by the partial charges of the MM atoms as an extra feature for the ML model that describes the QM region. This significantly reduces the computational time for training compared to methods like SO3krates \cite{frank2022so3krates}, since only a subset of atoms requires full periodically updated node features, while the rest can be captured through external interactions. While this strategy efficiently represents some long-range effects without sacrificing computational speed, it relies on a single scalar value (the electric field), which may not fully capture the complex nature of the interactions between MM and QM region.

One way to expand existing ML methods within the QM/MM methodology is via  the introduction of higher-order features through the multipole expansion \cite{edmonds1996angular,thorne1980multipole}.
The multipole expansion offers a systematic way to decompose electrostatic interactions into hierarchical terms, which naturally fits with equivariant neural networks, especially those based on spherical harmonics. In this work, we incorporate these features into the message-passing atomic cluster expansion (MACE) \cite{batatia2022mace} architecture, resulting in FieldMACE, and demonstrate how this approach can be applied to both ground- and excited-state simulations. Additionally, transfer learning from the foundational MACE-OFF \cite{kovacs2023mace} model originally trained on ground-state data without the inclusion of an environment is shown to significantly reduce the amount of data needed \cite{cai2020transfer,buterez2024transfer} for accurate predictions. Most notably, we show that this refinement can replicate reference population curves obtained from nonadiabatic excited-state dynamics \cite{mai2018nonadiabatic, Nelson2020}. These simulations are, when conducted with quantum chemistry, computationally intensive, often requiring months of computation\cite{Westermayr2022}, and still pose challenges to ML models \cite{Westermayr2021CR,mausenberger2024spainn,Tiefenbacher2025arXiv}. In this work, as little as 30 excited-state data points are required when using transfer learning for reproducing excited-state nonadiabatic molecular dynamics.

\section{Preliminaries and Related Work}

\subsection{Message-passing}
\label{sec:MP}

Atoms in a molecular system are characterized by their position vector $\mathbf{r}_i \in \mathbb{R}^3$ and atomic number $Z_i$.
These systems can be represented as graphs, where atoms correspond to nodes $\mathcal{V}_i$, and all nodes in the neighborhood of central atom $i$, $\mathcal{N}(i) = \{\mathcal{V}_j\ |\ ||\mathbf{r}_{ij}||_2 = ||\mathbf{r}_i - \mathbf{r}_j||_2 < r_\mathrm{cut} \}$ are connected by edges, where $r_\mathrm{cut}$ is a predefined cutoff radius. %, the definition of $\mathcal{N}(i)$ will be shown later.  
The categorical atomic numbers are transformed into a learnable invariant embedding vector $X_i\in \mathbb{R}^k$. During the message-passing process \cite{gilmer2017neural, jorgensen2018neural} each atom $i$ is associated with a set of features $\mathbf{h}_i \in \mathbb{R}^k$, and each edge $(i, j)$ is assigned the vector between the two atoms $\mathbf{r}_{ij}$. % = \mathbf{r}_i - \mathbf{r}_j$. %The features $\mathbf{h}_i$ are updated through an iterative process. The first feature set $\mathbf{h}_{i}^0$ is initially assigned $X_i$; following this, the general process is shown below,
The features are initialized with the atomic embedding $\mathbf{h}_{i}^0 = X_i$ and then updated through an iterative process:
\begin{align}
\mathbf{m}_{ij}^{(t)} & = \phi_{\text{msg}}\left(\mathbf{h}_i^{(t)}, \mathbf{h}_j^{(t)}, \mathbf{r}_{ij}\right)\\
\mathbf{m}_i^{(t)} & = \sum_{j \in \mathcal{N}(i)} \mathbf{m}_{ij}^{(t)}\\
\mathbf{h}_i^{(t+1)} & = \phi_{\text{update}}\left(\mathbf{h}_i^{(t)}, \mathbf{m}_i^{(t)}\right)
\end{align}
This process involves three main steps: i) creation of the messages $\mathbf{m}_{ij}^{(t)}$ that are computed for each edge using a learnable function $\phi_{\text{msg}}$ and the vector $\mathbf{r}_{ij}$. 
ii) These messages are aggregated over the neighborhood of each node using a permutation-invariant operation, such as summation. %The nodes which are aggregated must be in the neighborhood of node $i$, denoted as $\mathcal{N}(i)$, consisting of all nodes $j$ connected to $i$ by an edge, typically two nodes are connected if they are within a predefined cutoff radius of each other. 
iii) Finally, node features $\mathbf{h}_i$ are updated using another learnable function $\phi_{\text{update}}$. Messages passed between atoms in the neighborhood $\mathcal{N}(i)$ can represent scalar or higher-order features depending on the approach chosen. An illustration of the message-passing process can be seen in Fig.~\ref{fig:method}a (short-range block). 

\subsection{Equivariance}

Vectorial properties of a molecular system are uniquely defined by the orientation of it in space, \textit{i.e.} if the entire system rotates, associated properties have to transform accordingly, which is known as equivariance \cite{esteves2020theoretical}. Formally the set of features $\mathbf{h}_i$ is called equivariant under a group $G$ if  
\begin{equation}
D(R)\mathbf{h}_i(\mathbf{r}_1,..., \mathbf{r}_n) = \mathbf{h}_i(R \circ  (\mathbf{r}_1,..., \mathbf{r}_n)) \ ,
\end{equation}
where $R \in G $ and $D(R)$ is the equivalent operation acting on $\mathbf{h}_i$. The main focus of this paper will be on the special orthogonal group $SO(3)$. In the case of $SO(3)$, these equivariant features can be broken down into smaller building blocks \cite{thomas2018tensor}, the spherical harmonic functions, $Y_{\ell m}(\theta, \phi)$ acting on the sphere, where $\ell$ represents the degree and $m$ corresponds to the order, $m \in \{-\ell, ..., \ell\}$. Under a rotation $R \in SO(3)$, the spherical harmonics transform as follows:
\begin{equation}
Y_{\ell m'} (\theta', \phi') = \sum_{m}  D^{\ell}_{m'm}(R) Y_{\ell m} (\theta, \phi),
\end{equation}
with $D^{\ell}_{m'm}(R)$ being the Wigner-D matrices. In most cases, the equivariant feature vectors are expressed as a series of vectors where each is associated to a certain spherical harmonic, known as channels. The equivariant features are indexed as $\mathbf{h}^{\ell}_{i,m,k}$ where $k$ is the channel index. The message-passing framework can be rewritten in terms of these spherical harmonics. In this case only two body features will be considered, describing only interactions between neighbors around atom $i$.
\begin{align}
\mathbf{m}^{L,(t)}_{M, i , j , k} & = R_{\ell_1 m_1}(r_{ij}) \, Y_{\ell_1 m_1}(\theta, \phi) \mathbf{h}^{\ell_2, (t)}_{j,m_2,k} \\
\mathbf{m}^{\ell,(t)}_{m,i,k} & = \sum_{L,M} C_{L,M }^{lm} \sum_{j \in \mathcal{N}(i)} \mathbf{m}^{L,(t)}_{M, i , j , k} \label{eq:message_construct} \\
\mathbf{h}^{\ell,(t+1)}_{i,m,k} & = w_{mkk}^l \mathbf{m}^{\ell,(t)}_{m,i,k} + \mathbf{h}^{\ell, (t)}_{i,m,k} \label{eq:node_construct}\\
& L = (\ell_1, \ell_2), \quad M = (m_1, m_2) \nonumber
\end{align}
The angles $\theta$ and $\phi$ correspond to the directionality of the unit vector of $\mathbf{r_{ij}}$ and $r_{ij}$ the distance between atoms $i$ and $j$. The values $w_{mkk'}^l$ correspond to learnable, linear transformations across the channel of each spherical harmonic, an additional non-linear function can be placed along with the linear transformation. In order to preserve the correct equivariance for each order of spherical harmonics we must include the standard Clebsch Gordan coefficients $C_{L,M }^{lm}$ \cite{smorodinskiui1972clebsch}. The equations above form the basic block to the short range interactions in the following section. These blocks can also be exchanged with MACE which is discussed in the Appendix~\ref{sec:MACE_Architecture}.

\subsection{Multipole Expansion}

The multipole expansion is a technique that decomposes complex electromagnetic fields into hierarchical terms. This approach is particularly useful for analyzing fields far from their sources, where higher-order contributions become negligible. These moments are constructed in terms of spherical harmonics allowing a direct integration into equivariant neural networks that already use spherical harmonics such as MACE. 

A scalar potential $\Phi(\mathbf{r})$ arising from a charge distribution $\rho(\mathbf{r}')$, where $\mathbf{r}$ is the field point and $\mathbf{r}'$ is the source point, is defined by Coulomb's law as  
\begin{equation}
    \Phi(\mathbf{r}) = \frac{1}{4\pi\varepsilon_0} \int \frac{\rho(\mathbf{r}')}{|\mathbf{r} - \mathbf{r}'|} \, \mathrm{d}\mathbf{r}' \ .
\end{equation}
The function $\frac{1}{|\mathbf{r} - \mathbf{r}'|}$ can be expressed as a sum of spherical harmonics when $r < r'$:  
\begin{equation}
    \frac{1}{|\mathbf{r} - \mathbf{r}'|} = \sum_{\ell=0}^\infty \sum_{m=-\ell}^{\ell} \frac{4\pi}{2\ell + 1} \left( \frac{r}{r'} \right)^\ell Y_{\ell m}(\theta, \phi) {Y_{\ell m}}^*(\theta', \phi')
\end{equation}
%where $Y_{\ell m}(\theta, \phi)$ are spherical harmonics. 
Substituting this expansion into Coulomb’s law yields:  
\begin{equation}
    \Phi(\mathbf{r}) = \frac{1}{\varepsilon_0} \sum_{\ell=0}^\infty \sum_{m=-\ell}^\ell \frac{1}{2\ell + 1} \frac{Y_{\ell m}(\theta, \phi)}{r^{\ell + 1}} Q_{\ell m},
\end{equation}
where the multipole moment $Q_{\ell m}$ is defined as:  
\begin{equation}
    Q_{\ell m} = \int {r'}^\ell \rho(\mathbf{r}') {Y_{\ell m}}^*(\theta', \phi') \, \mathrm{d}\mathbf{r}'.
    \label{eq:multipole}
\end{equation}
Each term in this expansion corresponds to a specific multipole order, with higher-order terms decaying more rapidly with distance. These multipole moments will be used as feature vectors in the following section.

\section{Integration of QM/MM Framework with Equivariant Neural Networks}
\label{sec:MPE}

Our method splits a chemical system into a QM and an MM region. The descriptors for the QM region are derived in Sections~2.1 and~2.2. The MM atoms are described by their positions and partial charges that are assigned by a classical force field. While this is sufficient for the construction of the multipole expansion, the direct summation leads to significant information loss. To address this, we introduce weightings towards each of the MM atoms. Concretely, we compute pairwise attention coefficients \cite{vaswani2017attention} between the QM node features and the MM atom features to establish 
appropriate weightings.

\subsection{Attention Weightings}
Initial feature vectors for the MM atoms comprise spherical harmonics values, $Y^{\text{MM}}_{\ell m}$,  computed from the position values. These values are expanded to produce MM node features, 
\begin{align}
\mathbf{h}^{\ell, \text{MM}}_{j,m,k'} = w_{mk'}^{\ell} Y_{j,\ell m}^\text{MM}     
\end{align}
where $w_{mk'}^l$ is a mapping $\mathbb{R} \rightarrow \mathbb{R}^{k'}$.  To reduce computational costs, k' can be tuned as a hyperparameter and remains k'< k. A similar contraction operation is performed for the QM node features:
\begin{align}
\mathbf{h}^{\ell, \text{QM},(t)}_{i,m,k'} = w_{mk'k}^{\ell} \mathbf{h}^{\ell, \text{QM},(t)}_{i,m,k},    
\end{align}
where $w_{mk'k}^l$ is a mapping $\mathbb{R}^k \rightarrow \mathbb{R}^{k'}$. Given these feature vectors, pairwise attention is taken between QM atoms and each of the MM atoms. Note that no attention is computed between MM atoms. This significantly reduces the computational cost for large systems, full element-wise comparison scales with $O(n^2)$ whereas element-wise comparison between only MM and QM scales with $O(n)$ with the QM region being assumed to be constant. The element-wise comparisons are performed using an $L_1$ norm
\begin{align}
\mathbf{\kappa}_{ijk'}^{(t)} &= \xi_{mk'k'}^{\ell}(\lvert \mathbf{h}^{\ell,\text{QM},(t)}_{i,m,k'} -\mathbf{h}^{\ell, \text{MM}}_{j,m,k'}  \rvert)
\end{align}
Here $\xi$ consists of a linear transformation followed by a softmax normalization. These weighting values can be integrated into the multipole expansion to obtain the attention weighted expansion. An illustration of computing the attention weights can be seen beneath the long-range interaction block in Figure~\ref{fig:method}.

\subsection{Attention Multipole Expansion}

The charge distribution surrounding the QM region can be described by a set of point charges obtained from the MM atoms, $q_j^\text{MM}$;
\begin{align}
\rho(\mathbf{r}') &= 
\begin{cases}
q_j^\text{MM}, & \text{if } \mathbf{r}' = \mathbf{r}^\text{MM}_j,\\
0,   & \text{otherwise}.
\end{cases}
\end{align}
This simplifies the multipole moments, eq.~(\ref{eq:multipole}), into a sum over these discrete charges.%; in this case we do not use the spherical harmonic values but the MM node features leading to,
\begin{equation} Q^{\ell}_{m,k'} = \sum_{j} ({r^\text{MM}_j})^l q^\text{MM}_j \mathbf{h}^{\ell, \text{MM}}_{j,m,k'}  \ .
\end{equation}
It is important to note that the system is shifted so that the QM region is centered on the origin to maintain translational invariance. The multipole expansion can now be written including the weighting values for atom $i$ in the QM region,
\begin{align}
\Phi_{i,k'}^{\ell m,(t)}({\mathbf{r}^\text{QM}_i}) &= \frac{1}{\varepsilon_0} \sum_{j} \mathbf{\kappa}_{ijk'}^{(t)} \frac{Q^{\ell}_{m,j,k'}}{2\ell + 1} \frac{\mathbf{h}^{\ell, \text{QM},(t)}_{i,m,k'} }{|\mathbf{r}^\text{QM}_i - \mathbf{r}^\text{MM}_j|^{\ell + 1}} \\
Q^{\ell}_{m,j,k'} &=  ({r^\text{MM}_j})^l q^\text{MM}_j \mathbf{h}^{\ell, \text{MM}}_{j,m,k'} 
\end{align}
The term $\Phi_{i,k'}^{\ell m,(t)}(\mathbf{r}^\text{QM}_i)$ incorporates the necessary long-range information from the MM atoms for node feature $\mathbf{h}_i^{\text{QM},(t)}$ at atom $i$. 

\subsection{Message Construction}

The message is updated again after the short range message update eq.~\ref{eq:message_construct} by incorporating the multipole expansion term after expanding it to the correct dimensionality:
\begin{equation} 
\mathbf{m}^{\ell,(t)}_{m,i,k} = \mathbf{m}^{\ell,(t)}_{m,i,k} + w_{mkk'}^{\ell,(t)}\Phi_{i,k'}^{\ell m,(t)}({\mathbf{r}^\text{QM}_i})
\end{equation}
The node features are constructed from these messages in the same way as seen in eq.~(\ref{eq:node_construct}). These node features can be expanded to include many body interactions as well as other effects in the QM region. An illustration of the long range interaction implementation can be seen in the right half of Figure~\ref{fig:method}. Our long-range multipole message is integrated into the MACE framework.

\section{Experiments}
\begin{table*}[ht]
\centering
\caption{Mean Absolute Error (MAE) for Energy and Forces of FieldSchNet and FieldMACE models evaluated on the ground states of benzene, uracil, and retinoic acid systems. Results are shown for varying representation sizes (8, 16, 32, 64, 128) to assess model performance.}
\label{tab:FieldSchNet_FieldMACE_representation}
\begin{tabular}{llcccccc|ccccc}
\toprule
\multirow{2}{*}{\textbf{Model}} & \multirow{2}{*}{\textbf{System}} 
& \multicolumn{6}{c|}{\textbf{Energy (eV)}} 
& \multicolumn{5}{c}{\textbf{Force (eV/\AA)}} \\
\cmidrule(lr){3-8} \cmidrule(lr){9-13}
 &  & \multicolumn{5}{c}{\textbf{Representation Size}} &  
 & \multicolumn{5}{c}{\textbf{Representation Size}} \\
\cmidrule(lr){3-7} \cmidrule(lr){9-13}
 &  & \textbf{8} & \textbf{16} & \textbf{32} & \textbf{64} & \textbf{128} 
 &  & \textbf{8} & \textbf{16} & \textbf{32} & \textbf{64} & \textbf{128} \\
\midrule
\multirow{3}{*}{FieldSchNet} 
& Benzene       & 0.065 & 0.058 & 0.049 & 0.014 & 0.005 & 
 & 0.054 & 0.053 & 0.047 & 0.034 & 0.017 \\
& Uracil        & 0.182 & 0.151 & 0.079 & 0.016 & 0.015 & 
 & 0.128 & 0.118 & 0.112 & 0.054 & 0.048 \\
& Retinoic acid     & 0.145    & 0.053    & 0.029       & 0.028    & 0.015    & 
 & 0.190    & 0.141    & 0.099       & 0.064    & 0.031    \\
\midrule
\multirow{3}{*}{FieldMACE} 
& Benzene       & 0.041 & 0.033 & 0.018 & 0.006 & 0.003 & 
 & 0.039 & 0.038 & 0.031 & 0.018 & 0.010 \\
& Uracil        & 0.060 & 0.035 & 0.012 & 0.006 & 0.005 & 
 & 0.082 & 0.066 & 0.040 & 0.026 & 0.020 \\
& retinoic acid     & 0.061  & 0.036  & 0.019  & 0.013  & 0.012  & 
 & 0.117 & 0.092  & 0.065 & 0.043 & 0.028 \\
\bottomrule
\end{tabular}
\end{table*}

In this section, we evaluate the performance of our approach, called FieldMACE, which constitutes the inclusion of long-range message-passing into the MACE architecture.  
To this end, we test our model on the datasets by \cite{booeselt2021machine} comprising three molecular systems, namely benzene, uracil, and retinoic acid modeled under explicit environment. 
Each data set comprises molecules solvated in water, full details of the dataset can be found in the Appendix~\ref{sec:GS_datasets} or respective publication. We showcase the performance of our method using different representation sizes and compare it to another state-of-the-art model (FieldSchNet) that can be used for ML/MM, where ML is used to approximate the QM part of the system. In addition, we compute the vibrational power spectrum with FieldMACE for benzene solvated in water and compare the spectrum to a quantum chemical reference spectrum. In addition, we gauge FieldMACE’s capacity to handle electronically excited states by computing population curves resulting from nonadiabatic photodynamics simulations. Finally, we explore the potential for reusing parameters from foundational ground-state models trained without an external environment, particularly the MACE-OFF model,  which allows us to obtain lower errors with less amount of data compared to directly trained models.

\subsection{Ground State Simulations}

To evaluate the performance of FieldMACE in predicting energies and forces in the presence of an environment, we benchmark it against FieldSchNet \cite{Tiefenbacher2025arXiv}, a recently developed model that can be used for QM/MM simulations. Full architecture and training details can be found in the Appendix~C.  
To the best of our knowledge, there are currently few ML approaches capable of effectively describing molecular systems \cite{Hofstetter2022, Song2025,Tiefenbacher2025arXiv, yao2018tensormol, lier2022burnn, hofstetter2022graph, zinovjev2023electrostatic, thurlemann2023hybrid, zinovjev2024emle, semelak2024ani, grassano2024assessment, lei2024learning} in the presence of implicit or explicit environments. Many other models that are used for ML/MM or similar simulations often do not take the influence of an environment into account, by relying for instance on so-called mechanical embedding approaches (neglecting any electrostatic interactions), or by using specific embedding schemes, like the Buffered Region Neural Network approach (BuRNN),\cite{Lier2022JPCL} where the whole system is split into three regions rather than two and the ML model learns the difference between regions rather than the QM region. Thus, to allow for a fair comparison, the benchmark focuses exclusively on the FieldSchNet and FieldMACE models. The results, presented in Table~\ref{tab:FieldSchNet_FieldMACE_representation} summarize the mean absolute errors (MAEs) for energy and force predictions. %Table~\ref{tab:FieldSchNet_FieldMACE_representation} presents a comparison of the mean absolute error (MAE) for energy and force predictions between FieldSchNet and FieldMACE. 
As can be seen, FieldMACE consistently outperforms FieldSchNet, particularly at smaller representation sizes, demonstrating its capability to capture both short- and long-range interactions while being more compact. For energy predictions, both models show reduced errors with increasing representation size. %, with FieldMACE achieving the lowest MAE of 0.003~eV for benzene, compared to FieldSchNet that has a MAE of 0.005~eV. Similarly, for force predictions, FieldMACE achieves a superior minimum MAE of 0.010~eV/\AA~for benzene, outperforming FieldSchNet by 0.007~eV/\AA. 
Across all systems, FieldMACE consistently exhibits lower errors for both energy and forces. 

\begin{figure*}[htbp] % Use figure* for a
    \centering
    \includegraphics[width=\textwidth]{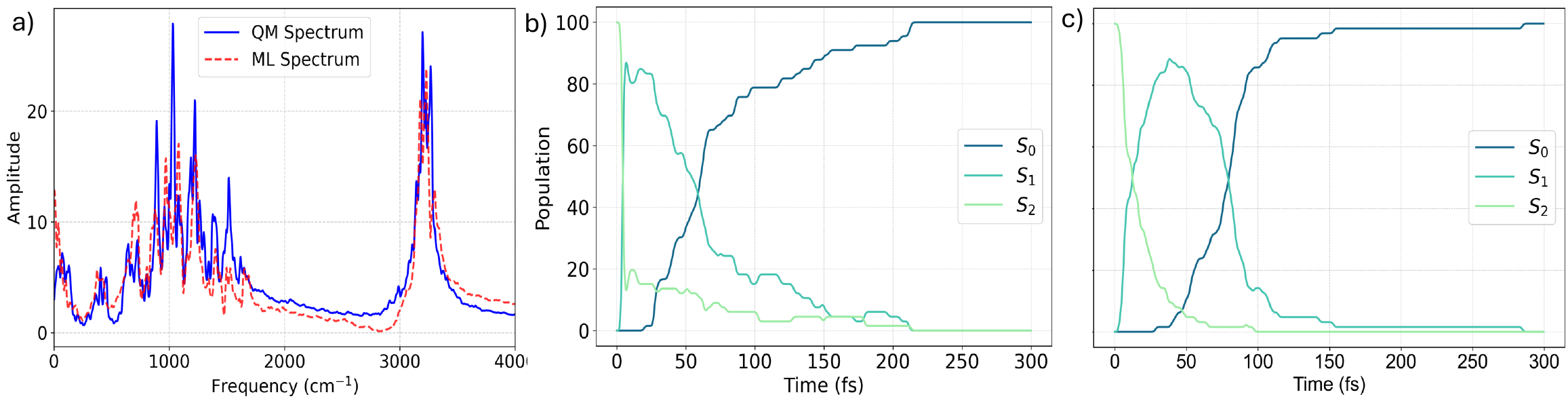} % Adjust the width 
    \caption{a) Power spectrum produced from the dynamics of benzene in a water box. Population curves resulting from photodynamics simulations using b) FieldMACE and c) quantum chemistry, showing the ring-opening reaction of furan after being excited to the second excited state ($S_2$). Transitions to $S_1$ and $S_0$ are present in both approaches. }
    \label{fig:population_dynamics}
\end{figure*}

In addition to evaluating test scores, we simulate the photodynamics of solvated benzene and compare the associated power spectra of the ML/MM and QM/MM reference dynamics in Figure~\ref{fig:population_dynamics}a, 
%shows the resulting power spectrum of benzene computed using FieldMACE and the QM reference, method and simulation 
computational details can be found in Appendix~B.2. The trained FieldMACE method produces a power spectrum in agreement with the QM spectrum, particularly, it reproduces the characteristic C–H stretching modes located near 3000~cm$^{-1}$ as well as the C=C ring stretching and C-H bending modes contained in the distributions centered around 1000~cm$^{-1}$.
In general, the peaks of the ML spectrum are located at the same frequencies as in the QM reference, only their relative intensity varies slightly. %The spectra are likely not fully converged, as the environment increases correlation times.

\subsection{Excited State Dynamics}

Going beyond ground state simulations, FieldMACE is used to simulate nonadiabatic dynamics using trajectory surface hopping (TSH) \cite{Tully1990,Tully1991IJQC}. These simulations describe the behavior of a molecule after being excited by light and are computationally much more demanding than ground-state simulations, especially when an explicit environment is included, as is the case here.
TSH recovers the quantum nature of the excited wave packet by averaging over many trajectories that have different initial conditions  (\textit{i.e.} configurations and momenta).
In Tully's Fewest Switches TSH \cite{Tully1990, Granucci2007}, used here, each trajectory changes its active state via so-called hops stochastically. For the photodynamics simulations, we integrated our FieldMACE approach into the SHARC (Surface Hopping including Arbitrary Couplings) program suite \cite{Richter2011sharc,SHARC3,mai2015general,mai2018nonadiabatic}. 

We investigate the molecule furan solvated in water to assess Field-MACE for excited-state dynamics. 
Details on the model and training are specified in Appendix~D. Upon excitation, the molecule can undergo a ring-opening reaction.
%Furan is a molecule with a five-membered ring containing an oxygen atom, the key feature in this reaction is the ring-opening reaction around the oxygen atom. 
We compare our model to recent simulations of this system using FieldSchNarc \cite{Tiefenbacher2025arXiv}.

Figure~\ref{fig:population_dynamics}b and c show the population dynamics obtained using FieldMACE (panel b) and the quantum chemical reference method (panel c). Population curves represent a distribution over the active state showing how the system evolves over time and hops between different electronic states. At the beginning of the simulations, the system is excited to the second excited singlet state, $S_2$. The nonadiabatic dynamics then show non-radiative decay back to the electronic ground state, S$_0$. 
FieldMACE accurately captures the rapid decay from the $S_2$ and the subsequent redistribution of populations to the $S_1$ and the ground state, $S_0$. The predicted population transfer closely matches quantum chemical reference populations. As can be seen, in both cases the $S_2$ population transfers quickly to the $S_1$ state, which is populated until around 100~fs, at which point the majority of the trajectories fall back into the ground state.

\section{Transferability}
\subsection{Ground State}

In this section, we analyze the possibility of reducing the amount of data necessary when starting with weights from pre-trained foundational ground-state models that have not been trained including an environment or excited-state effects.  Therefore, we leverage the foundational MACE model's pre-trained representations as initial parameters for FieldMACE's short-range modules to improve performance and data efficiency. %This study compared the transferability of FieldMACE when initialized with the foundational MACE model representation versus random initialization, focusing on its application to the previous datasets of retinoic acid, benzene, and uracil.
Full details on the models and training can be found in Appendix~B.

As can be seen in Figure ~\ref{fig:ground_transfer_results}, which shows learning curves using different amounts of data for fine-tuning for three molecular systems in solution, initializing the short-range blocks of 
FieldMACE with the foundational model weights reduces the amount of training data required to achieve comparable accuracy to models trained from scratch, particularly for force predictions (second line, panels d-f). %While the reduction in data needed for energy predictions was less substantial, it remained notable across all systems. 
Across all systems, the pre-trained initialization demonstrates a clear advantage, leading to improved accuracy compared to random initialization, even though the foundational models do not incorporate any solvent effects.

The benefits of using the foundational representation are most pronounced in systems with few strongly electronegative atoms, as is the case in benzene (first column, panels a and d) or retinoic acid (last column, panels c and f). These molecules are characterized by weaker intermolecular interactions and limited hydrogen bonding with the surrounding water molecules. The resulting potential energy surfaces for these systems will thus be more similar to the potential energy surfaces in a vacuum compared to the uracil system.

In contrast, for polar systems like uracil (middle column, panels b and e), which feature strong hydrogen bonding and complex electrostatic interactions, the benefits of the pre-trained initialization are less pronounced. While transfer-learned FieldMACE still outperforms the randomly initialized model, the benefit is smaller compared to the other examples. This suggests that the foundational model’s learned parameters have to be changed more in systems that interact more strongly with the MM environment. %are better suited for systems, where the interactions with the environment are less pronounced. 
Here, additional refinement on another dataset or task-specific training would improve performance for highly polar systems.

\begin{figure*}[htbp] % Use figure* for a wide figure
    \centering
    \includegraphics[width=\textwidth]{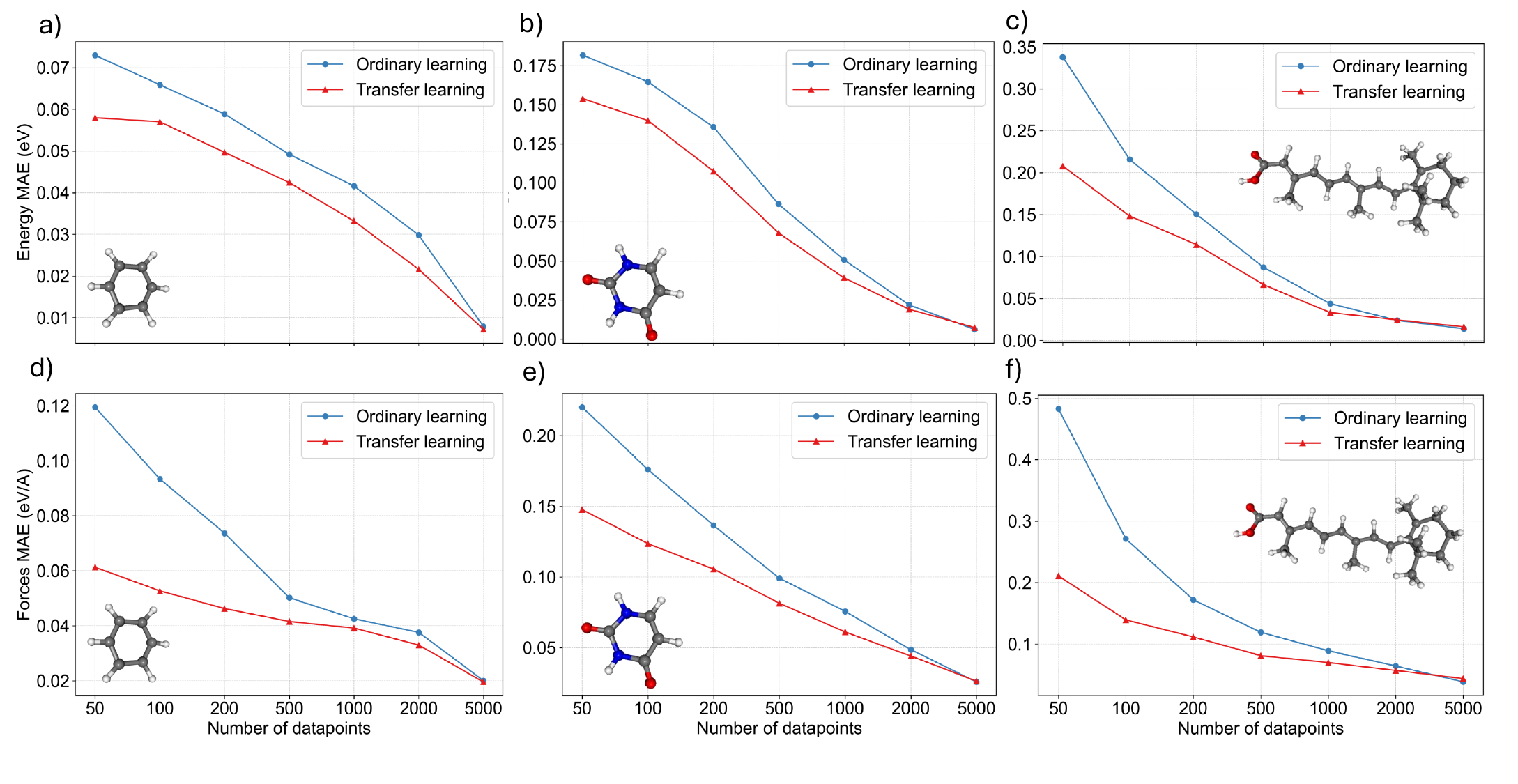}
    \caption{Comparison of transfer learning from MACE-OFF foundational model (red curves) against randomly initialized FieldMACE models (blue curves). Evaluation of  transfer and ordinary learning models with respect to energies (top row) and forces (bottom row) of molecules solvated in water: a) and d) benzene , b) and e) uracil and c) and f) retinoic acid . }
    \label{fig:ground_transfer_results}
\end{figure*}

\subsection{Excited States}

\begin{figure*}[htbp] % Use figurefor a wide figure
    \centering
    \includegraphics[scale=0.45]{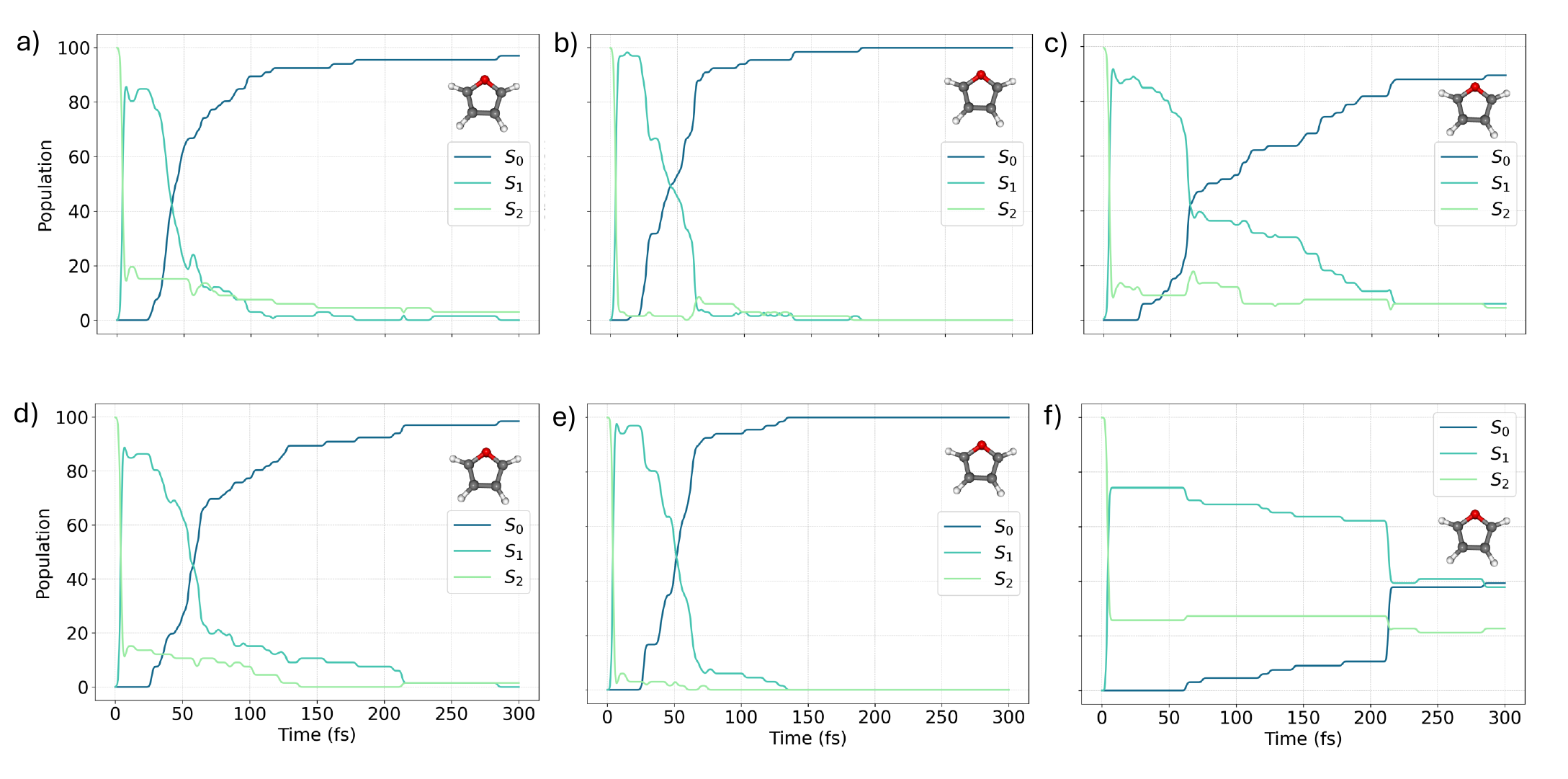}
    \caption{Population curves of furan starting in the third excited singlet state for transfer  learning models using a) 600, b) 150, and c) 30 data points and non-transfer learning models using d) 600, e) 150, and f) 30 data points.}
    \label{fig:excited_transfer_results}
\end{figure*}

%The transferability of FieldMACE in modeling excited states was evaluated using nonadiabatic dynamics simulations of furan as seen in the previous section. 
To further test the ability of FieldMACE to profit from transfer learning by heavily reducing the amount of training data needed, we revisit the furan in water system.  
Full details on the model and training can be found in Appendix~D. 
Rather than considering the mere test statistics (meaning error for energies and forces on the test set), we generate the population curves for the same set of initial conditions used in the previous section with FieldMACE models trained with a decreasing number of data. 
%Again the dynamics start from the second excited state, $S_2$, and the comparison is made for FieldMACE initialized with the foundational MACE model representation (transfer learning) and randomly initialized models (non-transfer learning) across datasets of 600, 150, and 30 datapoints. 
We compare FieldMACE models trained from scratch with randomly initialized parameters and those where the short-range block parameters are taken from a MACE-OFF model.
We use only 600, 150, and 30 data points to train these models.
Since excited-state data are computationally much more costly to obtain, this is a realistic use case.
The results are shown in Figure~\ref{fig:excited_transfer_results}.

For the largest dataset with 600 data points (first column), both transfer (panel a) and non-transfer models (panel d)  result in similar population curves that align well with the reference results (Figure \ref{fig:population_dynamics} c). Deviations in the $S_2$, $S_1$, and $S_0$ populations remain minimal throughout the simulation compared to the reference population curves. When the dataset size is reduced to 150 points (middle column), the predictions are reasonable for both transfer (panel b) and non-transfer (panel e) models. 

In contrast, for the smallest dataset of 30 data points (last column), a stark difference emerges between the two initialization strategies. The transfer model (panel c), leveraging the foundational representation, maintains a reasonable albeit worse approximation of the population curves and less stable simulations, capturing the general trends in decay and redistribution among electronic states. However, the model trained from scratch (panel f) fails to accurately capture the transition dynamics or the populations of $S_2$, $S_1$, and $S_0$. These results highlight the critical importance of transfer learning in low-data regimes, where the foundational representation enables FieldMACE to generalize effectively despite very limited training data.

Overall, the results demonstrate that transfer learning significantly enhances FieldMACE's performance, particularly when training data is scarce. With sufficient data (600 or 150 data points), both transfer and non-transfer models can perform reasonably well, though transfer learning provides an edge in accuracy. In low-data scenarios (30 data points), transfer learning is essential for maintaining reasonable predictions albeit less stable simulations, but the non-transfer model fails to capture the dynamics adequately. These findings emphasize the use of pre-trained foundational representations for obtaining accurate models.

\section{Discussion}

The attention weighted multipole expansion presented in this paper highlights an effective way of combining spherical harmomics-based equivariant neural networks with external charge distributions. In particular, results presented in this study highlight the significant advancements achieved by combining the multipole expansion with the MACE neural network architecture, FieldMACE, particularly its transferability and efficiency in modeling both ground-state and excited-state molecular processes. By leveraging the foundational MACE model representations, FieldMACE demonstrates its ability to extrapolate to new molecular systems while requiring less training data, especially for force predictions. This capability underscores the potential of pre-trained machine learning models in quantum chemistry.

One of the most notable findings is the effectiveness of FieldMACE in low-data regimes. For excited-state dynamics, the transfer learning approach from ground-state foundational models enabled FieldMACE to produce reasonable population dynamics even with as little as 30 data points, whereas models initialized with random weights failed to capture the underlying processes. This is particularly impactful given the computational expense associated with generating training data for excited-state simulations. The ability to maintain predictive accuracy with reduced datasets makes FieldMACE a practical choice for studying complex systems where data availability is limited.

When comparing FieldMACE with other molecular architectures that incorporate long-range interactions such as SO3Krates, the balance between accuracy and computational expense must be considered. In these architectures pairwise attention weightings are computed for all pairs in the system which leads to quadratic scaling, $O(n^2)$. In some cases considering interactions between all atoms in the system may be necessary but in many cases it is not. For instance, in the presented uracil system, which contains more than 3000 atoms, the only area of high significance is the uracil molecule. Training a single epoch while considering all pairwise interactions and assigning large node features can lead to prohibitively large computational costs to train these models. FieldMACE significantly reduces this computational burden by only considering interactions between the QM and MM atoms, thus allowing linear scaling when increasing the size of the external environment. %By avoiding all pairwise interactions in the whole system we obtain 
This leads to a dramatic reduction in computational overhead, and thus allows us to train on extended systems.

%Despite these successes, there are areas for improvement. 
%For highly polar systems and those dominated by strong hydrogen bonding, the foundational representation could be further enhanced by incorporating task-specific features or additional training on polar molecules. Similarly, 
%While FieldMACE demonstrates strong transferability for force predictions, the energy prediction accuracy, though reasonable, showed smaller gains compared to random initialization, suggesting room for improvement in this aspect.

In summary, FieldMACE represents a significant step forward in transferable machine learning models for molecular simulations in the presence of external charge distributions such as QM/MM simulations. Its ability to train on very large systems, combined with its accuracy in low-data regimes, makes it a powerful tool for QM/MM simulations in both ground and excited states. Future work will focus on enhancing its performance for highly polar systems, expanding its application to broader chemical spaces, and applying FieldMACE to systems such as proteins where long-range interactions are particularly dominant.

 \section*{Acknowledgements}
This work is funded in parts by the Deutsche Forschungsgemeinschaft (DFG) -- Project-ID 443871192 - GRK 2721: "Hydrogen Isotopes $^{1,2,3}$H". The authors acknowledge the ZIH TU Dresden, the URZ Leipzig University, and Paderborn Center for Parallel Computing (PC2) for providing the computational resources to conduct this study. 

\section*{Author Contributions}

RB (Conceptualization, Data curation, Formal analysis, Investigation, Methodology, Software, Visualization, Validation, Writing - original draft). 
JCBD (Methodology, Writing - review).
JW (Methodology, Conceptualization, Supervision, Writing - review).

\section*{Materials and Correspondence}
Correspondence to Julia Westermayr. 

\section*{Competing Interests}
The authors declare no competing interests.

\section*{Code and Data Availability}
The relevant code and datasets can be found here: \url{https://figshare.com/articles/dataset/Models_data_and_code_for_publication_Incorporating_Long-Range_Interactions_via_the_Multipole_Expansion_into_Ground_and_Excited-State_Molecular_Simulations_/28497857}
 and \url{https://github.com/rhyan10/FieldMACE/tree/master} , any additional datasets used can be accessed on request. 

\providecommand*{\mcitethebibliography}{\thebibliography}
\csname @ifundefined\endcsname{endmcitethebibliography}
{\let\endmcitethebibliography\endthebibliography}{}

\appendix
\section{MACE Architecture}
\label{sec:MACE_Architecture}
To extend message-passing neural networks (MPNNs) with many-body interactions, the atomic cluster expansion (ACE) \cite{drautz2019atomic} formulation can be utilized. This expansion incorporates higher-order terms, which are essential for accurately modeling complex molecular interactions. The message $\mathbf{m}_i^{(t)} $ at iteration t  is expressed as a sum over many-body contributions:

\begin{align*}
    \mathbf{m}_i^{(t)} &= \sum_{j \in \mathcal{N}_1(i)} \phi_1 \left( \mathbf{h}_i^{(t)}, \mathbf{h}_j^{(t)}, \mathbf{r}_{ij} \right) \\
    &+ \sum_{j_1 \in \mathcal{N}_2(i)} \sum_{j_2 \in \mathcal{N}_2(j_1)} \phi_2 \left( \mathbf{h}_i^{(t)}, \mathbf{h}_{j_1}^{(t)}, \mathbf{h}_{j_2}^{(t)}, \mathbf{r}_{ij_1}, \mathbf{r}_{j_1 j_2} \right) \\
    &\quad \vdots \\
    &+ \sum_{j_1, \dots, j_\nu \in \mathcal{N}_\nu(i)} \phi_\nu \left( \mathbf{h}_i^{(t)}, .., \mathbf{h}_{j_\nu}^{(t)}, \mathbf{r}_{ij_1}, .., \mathbf{r}_{j_{\nu-1} j_\nu} \right),
\end{align*}

where $N_\nu(i) $ denotes the set of $ \nu$-th order neighbors of node $ i $, and $ \phi_\nu $ represents the learnable interaction function incorporating many-body effects. This operation is computationally expensive but can be significantly reduced by constructing higher-order terms using tensor products of two body features then symmetrizing. The full derivations of the MACE architecture can be seen in the following sources \cite{batatia2025design, batatia2022mace}.

\section{Ground State Simulation Details}

\subsection{Molecular Datasets}
\label{sec:GS_datasets}

The configurations, energies, and forces for the three molecules solvated in water were taken from \citet{booeselt2021machine}.
Simulation details are briefly summarized here: All simulations were performed using a QM/MM approach with the DFT functionals BP86 \cite{becke1988density} for retonic acid, $\omega$B97X-D3 \cite{becke1993new, grimme2011effect} for uracil, and B2-PLYP  \cite{grimme2006semiempirical} for benzene; the def2-TZVP basis set was used for all calculations. The resolution of identity (RI) approximation \cite{dunlap1979first}, Grimme’s dispersion correction with Becke–Johnson damping \cite{grimme2011effect}, and electrostatic embedding were used to account for long-range interactions. The MM region consists of water molecules modeled with the SPC/E water model \cite{berendsen1987missing}. The QM/MM Hamiltonian incorporated MM point charges within a cutoff radius of 0.6~nm for benzene and 1.4~nm for uracil and retinoic acid.

Each solute was solvated in a periodic water box, with sizes of 1.6~nm, 3.3~nm and 5.0~nm respectively for benzene, uracil and retonic acid. The number of MM partial charges included was approximately 1500–2000 for uracil and 2500 for retinoic acid, while benzene had a smaller number of MM partial charges of 200 due to the smaller box size. The initial structures were generated using the ATB server \cite{malde2011automated} and relaxed at 0~K with a gradient descent algorithm until the predicted energy change fell below $0.1$~kJ~mol$^{-1}$.

For the molecular dynamics simulations that were performed, a time step of 0.5~fs was used, with a Nose–Hoover thermostat \cite{evans1985nose} (0.1~ps coupling constant) and a weak-coupling barostat (0.5~ps coupling constant). Long-range electrostatics beyond 1.4~nm were included using a reaction-field method. The system was simulated for 20,000 steps at 400~K and 1~bar, with the first 10,000 steps discarded for equilibration and the remaining 10,000 were saved.

\subsection{Simulation Details for benzene in water}
The following parameters were used to generate the power spectrum in Fig.~\ref{fig:population_dynamics}a. In the QM/MM simulations, an electrostatic embedding was used. The benzene atoms were treated quantum mechanically, while the surrounding water molecules were treated classically with the TIP3P \cite{mark2001structure} model. The simulation was done with the atomic simulation environment (ASE) \cite{larsen2017atomic}. The simulation cell was defined as a box of 16.0~Å per side with periodic boundary conditions applied in all directions. The QM region, was evaluated at a B3LYP \cite{becke1988density}/def-TZVP level.
For the ML/MM simulations of the same system, we used the FieldMACE model trained on the benzene dataset. The model used 128 channels with spherical harmonics up to degree 3. The vectors assigned to the MM atoms used spherical harmonics up to degree 3 and with 16 channels.

The initial geometry was optimized using the BFGS \cite{larsen2017atomic} minimizer until the maximum residual force was reduced to 0.1~eV/A, this was performed in the atomic simulation environment in ASE \cite{larsen2017atomic}. Initial velocities were drawn from a Maxwell-Boltzmann distribution at 300~K. We performed molecular dynamics with a Langevin integrator using a 1~fs timestep, a friction coefficient of 10~ps$^{-1}$, and a target temperature of 300~K.% to maintain appropriate thermal conditions. 
The system was propagated for a total of 15,000 steps. 

In order to facilitate a fair comparison between the ML and QM methods, the benzene data set from \cite{booeselt2021machine} was recomputed at the same level of theory, B3LYP-D3(BJ)/def-TZVP for the QM region with the TIP3P model for the MM region.
The ML model was trained on the recomputed frames from this data set (see details for the training in the following Sections).
The ML/MM-MD producing the ML-based power spectrum was performed with identical settings as the QM/MM-MD.

\section{Training and Model Details}

In all systems, a 5~{\AA}ngstrom cutoff was applied to the ML region to construct the local graph. The training parameters shown in the FieldMACE section below were also used for the transfer learning models, with the exception that the channel size was fixed to 128 for the transfer learning \cite{heydari2023transferring, buterez2024transfer} and the short range MACE blocks were initialized using the medium size foundational MACE-OFF model parameters for the transfer learning. Additionally, the other blocks such as the long range blocks and the readouts were randomly initialized. 

\subsection{FieldMACE}

All trainings were performed on a single NVIDIA H100 GPU. % with varying training times depending on the representation size used. 
For the three datasets (see Section~B.1.) we used all available configurations, which correspond to 10000 frames for benzene and uracil, and 9719 structures for retinoic acid. 
In all cases, we performed an 80:10:10 train:validation:test split.
%10\% of the dataset was used for validation and 90\% was used for training leading to 9000 training and 1000 validation data points for benzene and uracil; 8747 training and 972 validation datapoints for retonic acid respectively. 
For all FieldMACE models, we used two sets of short- and long-range layers. 
We trained models with varying channel sizes of 8, 16, 32, 64, 128.%, as shown in the table in the main paper. 
Additionally, all models were trained with spherical harmonics up to degree 3 and 8 Bessel basis functions. 
The initial vectors for the MM atoms consisted of spherical harmonics up to degree 3, combined with an embedding vector of size 16 (not varied between models). 

All models were trained with the loss function:
\begin{equation}
L = c_E \sum_{i=1}^n \bigl(E_{REF}^{(i)} - \hat{E}^{(i)}\bigl)^2 + c_F \sum_{i=1}^n \bigl(\mathbf{F}_{REF}^{(i)} - \hat{\mathbf{F}}^{(i)}\bigl)^2
\end{equation}
The coefficients $c_E$ and $c_F$ were both set to  100. The optimization was performed using the default optimizer in the MACE library  with an initial learning rate of 0.001. 

%\section{ Training Details}
\subsection{FieldSchNet}

All models used 6 interaction layers with 8-128 channels. A total of 25 Gaussian basis functions were used to featurize each edge. The learning rate was set to $10^{-4}$, with a decay factor of 0.8 triggered after 25 training epochs without improvement. Both energy and force terms were assigned a weight of 1.0 in the loss function, in order to get equal emphasis during optimization. Training took place on a single A30 GPU. The training splits used were identical to FieldMACE. %, with the total computational cost scaling according to the size of the dataset. 

\section{Excited State Simulations}
\subsection{Surface Hopping Molecular Dynamics}
Trajectory surface hopping (TSH) \cite{Tully1990,Tully1991IJQC,Richter2011sharc,mai2015general} is a nonadiabatic molecular dynamics approach where for each time step nuclear motion evolves classically on a single potential energy surface (PES), while stochastic hops between electronic states occur based on nonadiabatic coupling probabilities. In Tully’s fewest switches surface hopping (FSSH) \cite{Tully1990,Tully1991IJQC}, electronic state populations are propagated via the time-dependent Schrödinger equation,  
\begin{equation}
    \frac{d}{dt} c_k(t) = -\sum_{\ell} \left[ \frac{i}{\hbar} H_{k\ell} + \mathbf{d}_{k\ell} \cdot \mathbf{v} \right] c_\ell(t),
\end{equation}
where $c_k(t)$ denotes the time-dependent amplitude of electronic state $k$, $H_{k\ell}$ is the electronic Hamiltonian matrix element of states $k$ and $\ell$, $\mathbf{d}_{k\ell} = \langle \phi_k | \nabla_R | \phi_\ell \rangle$ is the nonadiabatic coupling vector between states $k$ and $\ell$, and $\mathbf{v}$ is the nuclear velocity vector. The probability of a hop from state $k$ to state $\ell$ in a time step $\Delta t$ is computed according to  
\begin{equation}
    P_{k \to \ell} = \max \left( 0, \frac{2 \operatorname{Re} \left( c_n^* c_\ell \left[ \frac{i}{\hbar} H_{k\ell} + \mathbf{d}_{k\ell} \cdot \mathbf{v}  \right] \right) }{|c_k|^2} \Delta t \right).
\end{equation}
At every time step, a random variable $\xi$ is drawn from the interval $[0,1]$, and the system transitions from state $k$ to state $\ell$ if $\xi$ is less than $P_{k \to \ell}$ \cite{Tully1990}. Since calculating explicit nonadiabatic couplings $\mathbf{d}_{k\ell}$ is resource-intensive, alternative methods like curvature-driven surface hopping \cite{Shu2022} have emerged. The curvature-driven surface hopping was utilized in all surface hopping simulations in this study. These techniques enable efficient propagation of surface hopping trajectories while accurately incorporating nonadiabatic effects in simulations of excited-state molecular dynamics. 

\subsection{Dataset Details}

The excited-state data set contains calculations of furan solvated in water (Zenodo archive \url{10.5281/zenodo.14536036}) and was produced for an ML/MM study using FieldSchNet \cite{Tiefenbacher2025arXiv}.
Computational details are briefly summarized here: The furan molecule is solvated in a cubix box (side length 15~\AA) of TIP3P water (1365 molecules). % and described with GAFF2 parameters and AM1-BCC charges. 
Initial structures for surface hopping dynamics were obtained from classical MD simulations. %These simulations were performed using \textsc{AMBER2022} with a 2fs timestep, a Langevin thermostat (friction constant 2ps$ ^{-1}$), and the SHAKE algorithm for hydrogen-heavy atom bonds.
%Following setup, the system was minimized for 2000 steps (steepest descent), heated for 20ps to 300K, and equilibrated for 600ps in the NPT ensemble with a Berendsen barostat.
%A 2.4ns production run then followed under the same conditions, saving snapshots every 4ps (600 frames). 
These frames were split into two sets for subsequent excited-state QM/MM simulations. One set contains the initial conditions for trajectories used for model training and validation, the other was used to compare dynamics of the trained model to unseen QM/MM simulations. %Only sets structures which fell inside a specific energy range were considered. 
We followed the same scheme to train and evaluate our models.
For more details on the initial conditions generation see \cite{Tiefenbacher2025arXiv}.

\subsection{Simulation and Training Details}

All nonadiabatic simulations were conducted using an interface between FieldMACE and the SHARC engine \cite{SHARC3,Richter2011sharc}. For the transfer learning experiments, the short-range blocks were initialized using the medium-size foundational MACE-OFF model parameters. The other blocks, such as the long-range blocks and the readouts, were randomly initialized.
Hyperparameters listed below were kept consistent for both the full data set and the smaller subsets used in transfer learning. These smaller datasets were constructed by taking the relevant amount of data randomly from the full dataset. 
A 5~{\AA}gstrom cutoff radius was applied along with 8 equally spaced Bessel basis functions.
Only two interaction layers were used, with 128 channels and spherical harmonics up to degree 3.
The initial vectors for the MM atoms were constructed from spherical harmonics up to degree 3, combined with an embedding vector of up to size 16. 
The readout functions were modified to produce multiple outputs rather than a single node.
In this case 3 states were used.
%depending on the number of energy levels.

%The loss function was an average RSME between the energy and force values over all the states predicted.  
The loss function was defined as mean squared error over all states for a property (energy and forces)
\begin{align}
%L = \frac{c_E}{n_L}\sum_{i=1}^n \bigl(E_{REF} - E\bigl)^2 + \frac{c_F}{n_L} \sum_{i=1}^n \bigl(F_{REF} - F\bigl)^2
L = & \frac{c_E}{n_L} \sum_{j=1}^{n_L} \sum_{i=1}^n \bigl(E_{REF}^{(i,j)} - \hat{E}^{(i,j)}\bigl)^2 \nonumber \\
& + \frac{c_F}{n_L} \sum_{j=1}^{n_L} \sum_{i=1}^n \bigl(\mathbf{F}_{REF}^{(i,j)} - \hat{\mathbf{F}}^{(i,j)}\bigl)^2
\end{align}
where $c_E$ and $c_F$ were both set to 100. The parameter $n_L$ represents the number of energy states, here 3. For optimization we used the default optimizer in the MACE library with a learning rate of 0.001. Training was performed on a single NVIDIA H100 GPU. %, with training times varying according to the number of data points used. 
Full details of the parameters used for the SHARC simulations can be found in \cite{Tiefenbacher2025arXiv}.


\begin{mcitethebibliography}{78}
\providecommand*{\natexlab}[1]{#1}
\providecommand*{\mciteSetBstSublistMode}[1]{}
\providecommand*{\mciteSetBstMaxWidthForm}[2]{}
\providecommand*{\mciteBstWouldAddEndPuncttrue}
  {\def\EndOfBibitem{\unskip.}}
\providecommand*{\mciteBstWouldAddEndPunctfalse}
  {\let\EndOfBibitem\relax}
\providecommand*{\mciteSetBstMidEndSepPunct}[3]{}
\providecommand*{\mciteSetBstSublistLabelBeginEnd}[3]{}
\providecommand*{\EndOfBibitem}{}
\mciteSetBstSublistMode{f}
\mciteSetBstMaxWidthForm{subitem}
{(\emph{\alph{mcitesubitemcount}})}
\mciteSetBstSublistLabelBeginEnd{\mcitemaxwidthsubitemform\space}
{\relax}{\relax}

\bibitem[Sauceda \emph{et~al.}(2022)Sauceda, G{\'a}lvez-Gonz{\'a}lez, Chmiela,
  Paz-Borb{\'o}n, M{\"u}ller, and Tkatchenko]{sauceda2022bigdml}
H.~E. Sauceda, L.~E. G{\'a}lvez-Gonz{\'a}lez, S.~Chmiela, L.~O. Paz-Borb{\'o}n,
  K.-R. M{\"u}ller and A.~Tkatchenko, \emph{Nat. Commun.}, 2022, \textbf{13},
  3733\relax
\mciteBstWouldAddEndPuncttrue
\mciteSetBstMidEndSepPunct{\mcitedefaultmidpunct}
{\mcitedefaultendpunct}{\mcitedefaultseppunct}\relax
\EndOfBibitem
\bibitem[Deng \emph{et~al.}(2023)Deng, Zhong, Jun, Riebesell, Han, Bartel, and
  Ceder]{deng2023chgnet}
B.~Deng, P.~Zhong, K.~Jun, J.~Riebesell, K.~Han, C.~J. Bartel and G.~Ceder,
  \emph{Nat. Mach. Intell.}, 2023, \textbf{5}, 1031--1041\relax
\mciteBstWouldAddEndPuncttrue
\mciteSetBstMidEndSepPunct{\mcitedefaultmidpunct}
{\mcitedefaultendpunct}{\mcitedefaultseppunct}\relax
\EndOfBibitem
\bibitem[Gao \emph{et~al.}(2020)Gao, Ramezanghorbani, Isayev, Smith, and
  Roitberg]{gao2020torchani}
X.~Gao, F.~Ramezanghorbani, O.~Isayev, J.~S. Smith and A.~E. Roitberg, \emph{J.
  Chem. Inf. Model.}, 2020, \textbf{60}, 3408--3415\relax
\mciteBstWouldAddEndPuncttrue
\mciteSetBstMidEndSepPunct{\mcitedefaultmidpunct}
{\mcitedefaultendpunct}{\mcitedefaultseppunct}\relax
\EndOfBibitem
\bibitem[Ko \emph{et~al.}(2021)Ko, Finkler, Goedecker, and
  Behler]{ko2021fourth}
T.~W. Ko, J.~A. Finkler, S.~Goedecker and J.~Behler, \emph{Nat. Commun.}, 2021,
  \textbf{12}, 398\relax
\mciteBstWouldAddEndPuncttrue
\mciteSetBstMidEndSepPunct{\mcitedefaultmidpunct}
{\mcitedefaultendpunct}{\mcitedefaultseppunct}\relax
\EndOfBibitem
\bibitem[Anstine and Isayev(2023)]{anstine2023machine}
D.~M. Anstine and O.~Isayev, \emph{J. Phys. Chem. A}, 2023, \textbf{127},
  2417--2431\relax
\mciteBstWouldAddEndPuncttrue
\mciteSetBstMidEndSepPunct{\mcitedefaultmidpunct}
{\mcitedefaultendpunct}{\mcitedefaultseppunct}\relax
\EndOfBibitem
\bibitem[Reichardt(2007)]{Reichardt2007}
C.~Reichardt, \emph{Org. Process Res. Dev.}, 2007, \textbf{11}, 105--113\relax
\mciteBstWouldAddEndPuncttrue
\mciteSetBstMidEndSepPunct{\mcitedefaultmidpunct}
{\mcitedefaultendpunct}{\mcitedefaultseppunct}\relax
\EndOfBibitem
\bibitem[Reichardt and Welton(2011)]{Reichardt2011}
C.~Reichardt and T.~Welton, \emph{Solvents and Solvent Effects in Organic
  Chemistry}, John Wiley \& Sons, 4th edn., 2011\relax
\mciteBstWouldAddEndPuncttrue
\mciteSetBstMidEndSepPunct{\mcitedefaultmidpunct}
{\mcitedefaultendpunct}{\mcitedefaultseppunct}\relax
\EndOfBibitem
\bibitem[Go and Taketomi(1978)]{Go1978}
N.~Go and H.~Taketomi, \emph{Proc. Natl. Acad. Sci.}, 1978, \textbf{75},
  559--563\relax
\mciteBstWouldAddEndPuncttrue
\mciteSetBstMidEndSepPunct{\mcitedefaultmidpunct}
{\mcitedefaultendpunct}{\mcitedefaultseppunct}\relax
\EndOfBibitem
\bibitem[Sagui and Darden(1999)]{Sagui1999}
C.~Sagui and T.~A. Darden, \emph{Annu. Rev. Biophys. Biomol. Struct.}, 1999,
  \textbf{28}, 155--179\relax
\mciteBstWouldAddEndPuncttrue
\mciteSetBstMidEndSepPunct{\mcitedefaultmidpunct}
{\mcitedefaultendpunct}{\mcitedefaultseppunct}\relax
\EndOfBibitem
\bibitem[Ambrosetti \emph{et~al.}(2014)Ambrosetti, Ferri, DiStasio, and
  Tkatchenko]{Ambrosetti2014}
A.~Ambrosetti, N.~Ferri, R.~A. DiStasio and A.~Tkatchenko, \emph{J. Chem.
  Phys.}, 2014, \textbf{140}, 18A508\relax
\mciteBstWouldAddEndPuncttrue
\mciteSetBstMidEndSepPunct{\mcitedefaultmidpunct}
{\mcitedefaultendpunct}{\mcitedefaultseppunct}\relax
\EndOfBibitem
\bibitem[Kang \emph{et~al.}(2024)Kang\emph{et~al.}]{Kang2024}
K.~Kang \emph{et~al.}, \emph{arXiv preprint}, 2024\relax
\mciteBstWouldAddEndPuncttrue
\mciteSetBstMidEndSepPunct{\mcitedefaultmidpunct}
{\mcitedefaultendpunct}{\mcitedefaultseppunct}\relax
\EndOfBibitem
\bibitem[Batzner \emph{et~al.}(2022)Batzner, Musaelian, Sun, Geiger, Mailoa,
  Kornbluth, Molinari, Smidt, and Kozinsky]{batzner20223}
S.~Batzner, A.~Musaelian, L.~Sun, M.~Geiger, J.~P. Mailoa, M.~Kornbluth,
  N.~Molinari, T.~E. Smidt and B.~Kozinsky, \emph{Nat. Commun.}, 2022,
  \textbf{13}, 2453\relax
\mciteBstWouldAddEndPuncttrue
\mciteSetBstMidEndSepPunct{\mcitedefaultmidpunct}
{\mcitedefaultendpunct}{\mcitedefaultseppunct}\relax
\EndOfBibitem
\bibitem[Sch{\"u}tt \emph{et~al.}(2021)Sch{\"u}tt, Unke, and
  Gastegger]{schutt2021equivariant}
K.~Sch{\"u}tt, O.~Unke and M.~Gastegger, Proc. Int. Conf. Mach. Learn., 2021,
  pp. 9377--9388\relax
\mciteBstWouldAddEndPuncttrue
\mciteSetBstMidEndSepPunct{\mcitedefaultmidpunct}
{\mcitedefaultendpunct}{\mcitedefaultseppunct}\relax
\EndOfBibitem
\bibitem[Gromiha and Selvaraj(1999)]{Gromiha1999}
M.~M. Gromiha and S.~Selvaraj, \emph{Biophys. Chem.}, 1999, \textbf{77},
  49--68\relax
\mciteBstWouldAddEndPuncttrue
\mciteSetBstMidEndSepPunct{\mcitedefaultmidpunct}
{\mcitedefaultendpunct}{\mcitedefaultseppunct}\relax
\EndOfBibitem
\bibitem[Westermayr \emph{et~al.}(2022)Westermayr, Chaudhuri, Jeindl, Hofmann,
  and Maurer]{westermayr2022long}
J.~Westermayr, S.~Chaudhuri, A.~Jeindl, O.~T. Hofmann and R.~J. Maurer,
  \emph{Digit. Discov.}, 2022, \textbf{1}, 463--475\relax
\mciteBstWouldAddEndPuncttrue
\mciteSetBstMidEndSepPunct{\mcitedefaultmidpunct}
{\mcitedefaultendpunct}{\mcitedefaultseppunct}\relax
\EndOfBibitem
\bibitem[Kosmala \emph{et~al.}(2023)Kosmala, Gasteiger, Gao, and
  G{\"u}nnemann]{kosmala2023ewald}
A.~Kosmala, J.~Gasteiger, N.~Gao and S.~G{\"u}nnemann, Proc. Int. Conf. Mach.
  Learn., 2023, pp. 17544--17563\relax
\mciteBstWouldAddEndPuncttrue
\mciteSetBstMidEndSepPunct{\mcitedefaultmidpunct}
{\mcitedefaultendpunct}{\mcitedefaultseppunct}\relax
\EndOfBibitem
\bibitem[Loche \emph{et~al.}(2024)Loche, Huguenin-Dumittan, Honarmand, Xu,
  Rumiantsev, How, Langer, and Ceriotti]{loche2024fast}
P.~Loche, K.~K. Huguenin-Dumittan, M.~Honarmand, Q.~Xu, E.~Rumiantsev, W.~B.
  How, M.~F. Langer and M.~Ceriotti, \emph{arXiv preprint arXiv:2412.03281},
  2024\relax
\mciteBstWouldAddEndPuncttrue
\mciteSetBstMidEndSepPunct{\mcitedefaultmidpunct}
{\mcitedefaultendpunct}{\mcitedefaultseppunct}\relax
\EndOfBibitem
\bibitem[Anstine and Isayev(2023)]{Anstine2023}
D.~M. Anstine and O.~Isayev, \emph{J. Phys. Chem. A}, 2023, \textbf{127},
  2417--2431\relax
\mciteBstWouldAddEndPuncttrue
\mciteSetBstMidEndSepPunct{\mcitedefaultmidpunct}
{\mcitedefaultendpunct}{\mcitedefaultseppunct}\relax
\EndOfBibitem
\bibitem[Gao and Remsing(2022)]{Gao2022}
A.~Gao and R.~C. Remsing, \emph{Nat. Commun.}, 2022, \textbf{13}, 1572\relax
\mciteBstWouldAddEndPuncttrue
\mciteSetBstMidEndSepPunct{\mcitedefaultmidpunct}
{\mcitedefaultendpunct}{\mcitedefaultseppunct}\relax
\EndOfBibitem
\bibitem[Behler and Cs{\'a}nyi(2021)]{behler2021machine}
J.~Behler and G.~Cs{\'a}nyi, \emph{Eur. Phys. J. B}, 2021, \textbf{94},
  1--11\relax
\mciteBstWouldAddEndPuncttrue
\mciteSetBstMidEndSepPunct{\mcitedefaultmidpunct}
{\mcitedefaultendpunct}{\mcitedefaultseppunct}\relax
\EndOfBibitem
\bibitem[Chanussot \emph{et~al.}(2021)Chanussot, Das, Goyal, Lavril, Shuaibi,
  Riviere, Tran, Heras-Domingo, Ho, Hu,\emph{et~al.}]{chanussot2021open}
L.~Chanussot, A.~Das, S.~Goyal, T.~Lavril, M.~Shuaibi, M.~Riviere, K.~Tran,
  J.~Heras-Domingo, C.~Ho, W.~Hu \emph{et~al.}, \emph{ACS Catal.}, 2021,
  \textbf{11}, 6059--6072\relax
\mciteBstWouldAddEndPuncttrue
\mciteSetBstMidEndSepPunct{\mcitedefaultmidpunct}
{\mcitedefaultendpunct}{\mcitedefaultseppunct}\relax
\EndOfBibitem
\bibitem[Tzeliou \emph{et~al.}(2022)Tzeliou, Mermigki, and
  Tzeli]{tzeliou2022review}
C.~E. Tzeliou, M.~A. Mermigki and D.~Tzeli, \emph{Molecules}, 2022,
  \textbf{27}, 2660\relax
\mciteBstWouldAddEndPuncttrue
\mciteSetBstMidEndSepPunct{\mcitedefaultmidpunct}
{\mcitedefaultendpunct}{\mcitedefaultseppunct}\relax
\EndOfBibitem
\bibitem[Boereboom \emph{et~al.}(2018)Boereboom, Fleurat-Lessard, and
  Bulo]{boereboom2018explicit}
J.~M. Boereboom, P.~Fleurat-Lessard and R.~E. Bulo, \emph{J. Chem. Theory
  Comput.}, 2018, \textbf{14}, 1841--1852\relax
\mciteBstWouldAddEndPuncttrue
\mciteSetBstMidEndSepPunct{\mcitedefaultmidpunct}
{\mcitedefaultendpunct}{\mcitedefaultseppunct}\relax
\EndOfBibitem
\bibitem[Frank \emph{et~al.}(2022)Frank, Unke, and
  M{\"u}ller]{frank2022so3krates}
T.~Frank, O.~Unke and K.-R. M{\"u}ller, \emph{Adv. Neural Inf. Process. Syst.},
  2022, \textbf{35}, 29400--29413\relax
\mciteBstWouldAddEndPuncttrue
\mciteSetBstMidEndSepPunct{\mcitedefaultmidpunct}
{\mcitedefaultendpunct}{\mcitedefaultseppunct}\relax
\EndOfBibitem
\bibitem[Batatia \emph{et~al.}(2023)Batatia, Schaaf, Chen, Cs{\'a}nyi, Ortner,
  and Faber]{batatia2023equivariant}
I.~Batatia, L.~L. Schaaf, H.~Chen, G.~Cs{\'a}nyi, C.~Ortner and F.~A. Faber,
  \emph{arXiv preprint arXiv:2310.10434}, 2023\relax
\mciteBstWouldAddEndPuncttrue
\mciteSetBstMidEndSepPunct{\mcitedefaultmidpunct}
{\mcitedefaultendpunct}{\mcitedefaultseppunct}\relax
\EndOfBibitem
\bibitem[Frank \emph{et~al.}(2024)Frank\emph{et~al.}]{Frank2024}
J.~T. Frank \emph{et~al.}, \emph{arXiv preprint}, 2024\relax
\mciteBstWouldAddEndPuncttrue
\mciteSetBstMidEndSepPunct{\mcitedefaultmidpunct}
{\mcitedefaultendpunct}{\mcitedefaultseppunct}\relax
\EndOfBibitem
\bibitem[Gastegger \emph{et~al.}(2021)Gastegger, Sch{\"u}tt, and
  M{\"u}ller]{gastegger2021machine}
M.~Gastegger, K.~T. Sch{\"u}tt and K.-R. M{\"u}ller, \emph{Chem. Sci.}, 2021,
  \textbf{12}, 11473--11483\relax
\mciteBstWouldAddEndPuncttrue
\mciteSetBstMidEndSepPunct{\mcitedefaultmidpunct}
{\mcitedefaultendpunct}{\mcitedefaultseppunct}\relax
\EndOfBibitem
\bibitem[Edmonds(1996)]{edmonds1996angular}
A.~R. Edmonds, \emph{Angular momentum in quantum mechanics}, Princeton
  university press, 1996, vol.~4\relax
\mciteBstWouldAddEndPuncttrue
\mciteSetBstMidEndSepPunct{\mcitedefaultmidpunct}
{\mcitedefaultendpunct}{\mcitedefaultseppunct}\relax
\EndOfBibitem
\bibitem[Thorne(1980)]{thorne1980multipole}
K.~S. Thorne, \emph{Rev. Mod. Phys.}, 1980, \textbf{52}, 299\relax
\mciteBstWouldAddEndPuncttrue
\mciteSetBstMidEndSepPunct{\mcitedefaultmidpunct}
{\mcitedefaultendpunct}{\mcitedefaultseppunct}\relax
\EndOfBibitem
\bibitem[Batatia \emph{et~al.}(2022)Batatia, Kovacs, Simm, Ortner, and
  Cs{\'a}nyi]{batatia2022mace}
I.~Batatia, D.~P. Kovacs, G.~Simm, C.~Ortner and G.~Cs{\'a}nyi, \emph{Adv.
  Neural Inf. Process. Syst.}, 2022, \textbf{35}, 11423--11436\relax
\mciteBstWouldAddEndPuncttrue
\mciteSetBstMidEndSepPunct{\mcitedefaultmidpunct}
{\mcitedefaultendpunct}{\mcitedefaultseppunct}\relax
\EndOfBibitem
\bibitem[Kov{\'a}cs \emph{et~al.}(2023)Kov{\'a}cs, Moore, Browning, Batatia,
  Horton, Kapil, Witt, Magd{\u{a}}u, Cole, and Cs{\'a}nyi]{kovacs2023mace}
D.~P. Kov{\'a}cs, J.~H. Moore, N.~J. Browning, I.~Batatia, J.~T. Horton,
  V.~Kapil, W.~C. Witt, I.-B. Magd{\u{a}}u, D.~J. Cole and G.~Cs{\'a}nyi,
  \emph{arXiv preprint arXiv:2312.15211}, 2023\relax
\mciteBstWouldAddEndPuncttrue
\mciteSetBstMidEndSepPunct{\mcitedefaultmidpunct}
{\mcitedefaultendpunct}{\mcitedefaultseppunct}\relax
\EndOfBibitem
\bibitem[Cai \emph{et~al.}(2020)Cai, Wang, Xu, Zhang, Tang, Ouyang, Lai, and
  Pei]{cai2020transfer}
C.~Cai, S.~Wang, Y.~Xu, W.~Zhang, K.~Tang, Q.~Ouyang, L.~Lai and J.~Pei,
  \emph{J. Med. Chem.}, 2020, \textbf{63}, 8683--8694\relax
\mciteBstWouldAddEndPuncttrue
\mciteSetBstMidEndSepPunct{\mcitedefaultmidpunct}
{\mcitedefaultendpunct}{\mcitedefaultseppunct}\relax
\EndOfBibitem
\bibitem[Buterez \emph{et~al.}(2024)Buterez, Janet, Kiddle, Oglic, and
  Li{\'o}]{buterez2024transfer}
D.~Buterez, J.~P. Janet, S.~J. Kiddle, D.~Oglic and P.~Li{\'o}, \emph{Nat.
  Commun.}, 2024, \textbf{15}, 1517\relax
\mciteBstWouldAddEndPuncttrue
\mciteSetBstMidEndSepPunct{\mcitedefaultmidpunct}
{\mcitedefaultendpunct}{\mcitedefaultseppunct}\relax
\EndOfBibitem
\bibitem[Mai \emph{et~al.}(2018)Mai, Marquetand, and
  Gonz{\'a}lez]{mai2018nonadiabatic}
S.~Mai, P.~Marquetand and L.~Gonz{\'a}lez, \emph{Wiley Interdiscip. Rev.
  Comput. Mol. Sci.}, 2018, \textbf{8}, e1370\relax
\mciteBstWouldAddEndPuncttrue
\mciteSetBstMidEndSepPunct{\mcitedefaultmidpunct}
{\mcitedefaultendpunct}{\mcitedefaultseppunct}\relax
\EndOfBibitem
\bibitem[Nelson \emph{et~al.}(2020)Nelson, White, Bjorgaard, Sifain, Zhang,
  Nebgen, Fernandez-Alberti, Mozyrsky, Roitberg, and Tretiak]{Nelson2020}
T.~R. Nelson, A.~J. White, J.~A. Bjorgaard, A.~E. Sifain, Y.~Zhang, B.~T.
  Nebgen, S.~Fernandez-Alberti, D.~Mozyrsky, A.~E. Roitberg and S.~Tretiak,
  \emph{Chem. Rev.}, 2020, \textbf{120}, 2215--2287\relax
\mciteBstWouldAddEndPuncttrue
\mciteSetBstMidEndSepPunct{\mcitedefaultmidpunct}
{\mcitedefaultendpunct}{\mcitedefaultseppunct}\relax
\EndOfBibitem
\bibitem[Westermayr \emph{et~al.}(2022)Westermayr, Ghosh, Mori, and
  Gonz{\'a}lez]{Westermayr2022}
J.~Westermayr, S.~Ghosh, T.~Mori and L.~Gonz{\'a}lez, \emph{Nat. Chem.}, 2022,
  \textbf{14}, 914--919\relax
\mciteBstWouldAddEndPuncttrue
\mciteSetBstMidEndSepPunct{\mcitedefaultmidpunct}
{\mcitedefaultendpunct}{\mcitedefaultseppunct}\relax
\EndOfBibitem
\bibitem[Westermayr and Marquetand(2021)]{Westermayr2021CR}
J.~Westermayr and P.~Marquetand, \emph{Chem. Rev.}, 2021, \textbf{121},
  9873--9926\relax
\mciteBstWouldAddEndPuncttrue
\mciteSetBstMidEndSepPunct{\mcitedefaultmidpunct}
{\mcitedefaultendpunct}{\mcitedefaultseppunct}\relax
\EndOfBibitem
\bibitem[Mausenberger \emph{et~al.}(2024)Mausenberger, M{\"u}ller, Tkatchenko,
  Marquetand, Gonz{\'a}lez, and Westermayr]{mausenberger2024spainn}
S.~Mausenberger, C.~M{\"u}ller, A.~Tkatchenko, P.~Marquetand, L.~Gonz{\'a}lez
  and J.~Westermayr, \emph{Chem. Sci.}, 2024, \textbf{15}, 15880--15890\relax
\mciteBstWouldAddEndPuncttrue
\mciteSetBstMidEndSepPunct{\mcitedefaultmidpunct}
{\mcitedefaultendpunct}{\mcitedefaultseppunct}\relax
\EndOfBibitem
\bibitem[Tiefenbacher \emph{et~al.}(2025)Tiefenbacher, Bachmair, Chen,
  Westermayr, Marquetand, Dietschreit, and Gonz{\'a}lez]{Tiefenbacher2025arXiv}
M.~X. Tiefenbacher, B.~Bachmair, C.~G. Chen, J.~Westermayr, P.~Marquetand,
  J.~C. Dietschreit and L.~Gonz{\'a}lez, \emph{Excited-state nonadiabatic
  dynamics in explicit solvent using machine learned interatomic potentials},
  2025\relax
\mciteBstWouldAddEndPuncttrue
\mciteSetBstMidEndSepPunct{\mcitedefaultmidpunct}
{\mcitedefaultendpunct}{\mcitedefaultseppunct}\relax
\EndOfBibitem
\bibitem[Gilmer \emph{et~al.}(2017)Gilmer, Schoenholz, Riley, Vinyals, and
  Dahl]{gilmer2017neural}
J.~Gilmer, S.~S. Schoenholz, P.~F. Riley, O.~Vinyals and G.~E. Dahl, Int. Conf.
  Mach. Learn., 2017, pp. 1263--1272\relax
\mciteBstWouldAddEndPuncttrue
\mciteSetBstMidEndSepPunct{\mcitedefaultmidpunct}
{\mcitedefaultendpunct}{\mcitedefaultseppunct}\relax
\EndOfBibitem
\bibitem[J{\o}rgensen \emph{et~al.}(2018)J{\o}rgensen, Jacobsen, and
  Schmidt]{jorgensen2018neural}
P.~B. J{\o}rgensen, K.~W. Jacobsen and M.~N. Schmidt, \emph{arXiv preprint
  arXiv:1806.03146}, 2018\relax
\mciteBstWouldAddEndPuncttrue
\mciteSetBstMidEndSepPunct{\mcitedefaultmidpunct}
{\mcitedefaultendpunct}{\mcitedefaultseppunct}\relax
\EndOfBibitem
\bibitem[Esteves(2020)]{esteves2020theoretical}
C.~Esteves, \emph{arXiv preprint arXiv:2004.05154}, 2020\relax
\mciteBstWouldAddEndPuncttrue
\mciteSetBstMidEndSepPunct{\mcitedefaultmidpunct}
{\mcitedefaultendpunct}{\mcitedefaultseppunct}\relax
\EndOfBibitem
\bibitem[Thomas \emph{et~al.}(2018)Thomas, Smidt, Kearnes, Yang, Li, Kohlhoff,
  and Riley]{thomas2018tensor}
N.~Thomas, T.~Smidt, S.~Kearnes, L.~Yang, L.~Li, K.~Kohlhoff and P.~Riley,
  \emph{arXiv preprint arXiv:1802.08219}, 2018\relax
\mciteBstWouldAddEndPuncttrue
\mciteSetBstMidEndSepPunct{\mcitedefaultmidpunct}
{\mcitedefaultendpunct}{\mcitedefaultseppunct}\relax
\EndOfBibitem
\bibitem[Smorodinski{\u\i} and Shelepin(1972)]{smorodinskiui1972clebsch}
Y.~A. Smorodinski{\u\i} and L.~A. Shelepin, \emph{Sov. Phys. Usp.}, 1972,
  \textbf{15}, 1\relax
\mciteBstWouldAddEndPuncttrue
\mciteSetBstMidEndSepPunct{\mcitedefaultmidpunct}
{\mcitedefaultendpunct}{\mcitedefaultseppunct}\relax
\EndOfBibitem
\bibitem[Vaswani(2017)]{vaswani2017attention}
A.~Vaswani, \emph{Adv. Neural Inf. Process. Syst.}, 2017\relax
\mciteBstWouldAddEndPuncttrue
\mciteSetBstMidEndSepPunct{\mcitedefaultmidpunct}
{\mcitedefaultendpunct}{\mcitedefaultseppunct}\relax
\EndOfBibitem
\bibitem[Bo\"oselt \emph{et~al.}(2021)Bo\"oselt, Th\"urlemann, and
  Riniker]{booeselt2021machine}
L.~Bo\"oselt, M.~Th\"urlemann and S.~Riniker, \emph{J. Chem. Theory Comput.},
  2021, \textbf{17}, 2641--2658\relax
\mciteBstWouldAddEndPuncttrue
\mciteSetBstMidEndSepPunct{\mcitedefaultmidpunct}
{\mcitedefaultendpunct}{\mcitedefaultseppunct}\relax
\EndOfBibitem
\bibitem[Hofstetter \emph{et~al.}(2022)Hofstetter, B{\"o}selt, and
  Riniker]{Hofstetter2022}
A.~Hofstetter, L.~B{\"o}selt and S.~Riniker, \emph{Phys. Chem. Chem. Phys.},
  2022, \textbf{24}, 22497--22512\relax
\mciteBstWouldAddEndPuncttrue
\mciteSetBstMidEndSepPunct{\mcitedefaultmidpunct}
{\mcitedefaultendpunct}{\mcitedefaultseppunct}\relax
\EndOfBibitem
\bibitem[Song and Yang(2025)]{Song2025}
G.~Song and W.~Yang, \emph{arXiv preprint}, 2025\relax
\mciteBstWouldAddEndPuncttrue
\mciteSetBstMidEndSepPunct{\mcitedefaultmidpunct}
{\mcitedefaultendpunct}{\mcitedefaultseppunct}\relax
\EndOfBibitem
\bibitem[Yao \emph{et~al.}(2018)Yao, Herr, Toth, Mckintyre, and
  Parkhill]{yao2018tensormol}
K.~Yao, J.~E. Herr, D.~W. Toth, R.~Mckintyre and J.~Parkhill, \emph{Chem.
  Sci.}, 2018, \textbf{9}, 2261--2269\relax
\mciteBstWouldAddEndPuncttrue
\mciteSetBstMidEndSepPunct{\mcitedefaultmidpunct}
{\mcitedefaultendpunct}{\mcitedefaultseppunct}\relax
\EndOfBibitem
\bibitem[Lier \emph{et~al.}(2022)Lier, Poliak, Marquetand, Westermayr, and
  Oostenbrink]{lier2022burnn}
B.~Lier, P.~Poliak, P.~Marquetand, J.~Westermayr and C.~Oostenbrink, \emph{J.
  Phys. Chem. Lett.}, 2022, \textbf{13}, 3812--3818\relax
\mciteBstWouldAddEndPuncttrue
\mciteSetBstMidEndSepPunct{\mcitedefaultmidpunct}
{\mcitedefaultendpunct}{\mcitedefaultseppunct}\relax
\EndOfBibitem
\bibitem[Hofstetter \emph{et~al.}(2022)Hofstetter, Böselt, and
  Riniker]{hofstetter2022graph}
A.~Hofstetter, L.~Böselt and S.~Riniker, \emph{Phys. Chem. Chem. Phys.}, 2022,
  \textbf{24}, 22497--22512\relax
\mciteBstWouldAddEndPuncttrue
\mciteSetBstMidEndSepPunct{\mcitedefaultmidpunct}
{\mcitedefaultendpunct}{\mcitedefaultseppunct}\relax
\EndOfBibitem
\bibitem[Zinovjev(2023)]{zinovjev2023electrostatic}
K.~Zinovjev, \emph{J. Chem. Theory Comput.}, 2023, \textbf{19},
  1888--1897\relax
\mciteBstWouldAddEndPuncttrue
\mciteSetBstMidEndSepPunct{\mcitedefaultmidpunct}
{\mcitedefaultendpunct}{\mcitedefaultseppunct}\relax
\EndOfBibitem
\bibitem[Thürlemann and Riniker(2023)]{thurlemann2023hybrid}
M.~Thürlemann and S.~Riniker, \emph{Chem. Sci.}, 2023, \textbf{14},
  12661--12675\relax
\mciteBstWouldAddEndPuncttrue
\mciteSetBstMidEndSepPunct{\mcitedefaultmidpunct}
{\mcitedefaultendpunct}{\mcitedefaultseppunct}\relax
\EndOfBibitem
\bibitem[Zinovjev \emph{et~al.}(2024)Zinovjev, Hedges, Montagud~Andreu, Woods,
  Tuñón, and van~der Kamp]{zinovjev2024emle}
K.~Zinovjev, L.~Hedges, R.~Montagud~Andreu, C.~Woods, I.~Tuñón and M.~W.
  van~der Kamp, \emph{J. Chem. Theory Comput.}, 2024\relax
\mciteBstWouldAddEndPuncttrue
\mciteSetBstMidEndSepPunct{\mcitedefaultmidpunct}
{\mcitedefaultendpunct}{\mcitedefaultseppunct}\relax
\EndOfBibitem
\bibitem[Semelak \emph{et~al.}(2024)Semelak, Pickering, Huddleston, Olmos,
  Grassano, Clemente, Drusin, Marti, González,
  Roitberg,\emph{et~al.}]{semelak2024ani}
J.~A. Semelak, I.~Pickering, K.~K. Huddleston, J.~Olmos, J.~S. Grassano,
  C.~Clemente, S.~I. Drusin, M.~Marti, M.~C.~G. González, A.~E. Roitberg
  \emph{et~al.}, 2024\relax
\mciteBstWouldAddEndPuncttrue
\mciteSetBstMidEndSepPunct{\mcitedefaultmidpunct}
{\mcitedefaultendpunct}{\mcitedefaultseppunct}\relax
\EndOfBibitem
\bibitem[Grassano \emph{et~al.}(2024)Grassano, Pickering, Roitberg,
  González~Lebrero, Estrin, and Semelak]{grassano2024assessment}
J.~S. Grassano, I.~Pickering, A.~E. Roitberg, M.~C. González~Lebrero, D.~A.
  Estrin and J.~A. Semelak, \emph{J. Chem. Inf. Model.}, 2024, \textbf{64},
  4047--4058\relax
\mciteBstWouldAddEndPuncttrue
\mciteSetBstMidEndSepPunct{\mcitedefaultmidpunct}
{\mcitedefaultendpunct}{\mcitedefaultseppunct}\relax
\EndOfBibitem
\bibitem[Lei \emph{et~al.}(2024)Lei, Yagi, and Sugita]{lei2024learning}
Y.-K. Lei, K.~Yagi and Y.~Sugita, \emph{J. Chem. Phys.}, 2024, \textbf{160},
  214109\relax
\mciteBstWouldAddEndPuncttrue
\mciteSetBstMidEndSepPunct{\mcitedefaultmidpunct}
{\mcitedefaultendpunct}{\mcitedefaultseppunct}\relax
\EndOfBibitem
\bibitem[Lier \emph{et~al.}(2022)Lier, Poliak, Marquetand, Westermayr, and
  Oostenbrink]{Lier2022JPCL}
B.~Lier, P.~Poliak, P.~Marquetand, J.~Westermayr and C.~Oostenbrink, \emph{J.
  Phys. Chem. Lett.}, 2022, \textbf{13}, 3812--3818\relax
\mciteBstWouldAddEndPuncttrue
\mciteSetBstMidEndSepPunct{\mcitedefaultmidpunct}
{\mcitedefaultendpunct}{\mcitedefaultseppunct}\relax
\EndOfBibitem
\bibitem[Tully(1990)]{Tully1990}
J.~C. Tully, \emph{J. Chem. Phys.}, 1990, \textbf{92}, 1061--1071\relax
\mciteBstWouldAddEndPuncttrue
\mciteSetBstMidEndSepPunct{\mcitedefaultmidpunct}
{\mcitedefaultendpunct}{\mcitedefaultseppunct}\relax
\EndOfBibitem
\bibitem[Tully(1991)]{Tully1991IJQC}
J.~C. Tully, \emph{Int. J. Quantum Chem.}, 1991, \textbf{40}, 299--309\relax
\mciteBstWouldAddEndPuncttrue
\mciteSetBstMidEndSepPunct{\mcitedefaultmidpunct}
{\mcitedefaultendpunct}{\mcitedefaultseppunct}\relax
\EndOfBibitem
\bibitem[Granucci and Persico(2007)]{Granucci2007}
G.~Granucci and M.~Persico, \emph{J. Chem. Phys.}, 2007, \textbf{126},
  134114\relax
\mciteBstWouldAddEndPuncttrue
\mciteSetBstMidEndSepPunct{\mcitedefaultmidpunct}
{\mcitedefaultendpunct}{\mcitedefaultseppunct}\relax
\EndOfBibitem
\bibitem[Richter \emph{et~al.}(2011)Richter, Marquetand,
  Gonz{\'a}lez-V\'{a}zquez, Sola, and Gonz{\'a}lez]{Richter2011sharc}
M.~Richter, P.~Marquetand, J.~Gonz{\'a}lez-V\'{a}zquez, I.~Sola and
  L.~Gonz{\'a}lez, \emph{J. Chem. Theory Comput.}, 2011, \textbf{7},
  1253--1258\relax
\mciteBstWouldAddEndPuncttrue
\mciteSetBstMidEndSepPunct{\mcitedefaultmidpunct}
{\mcitedefaultendpunct}{\mcitedefaultseppunct}\relax
\EndOfBibitem
\bibitem[Mai \emph{et~al.}(2023)Mai, Avagliano, Heindl, Marquetand, Menger,
  Oppel, Plasser, Polonius, Ruckenbauer, Shu, Truhlar, Zhang, Zobel, and
  Gonz{\'a}lez]{SHARC3}
S.~Mai, D.~Avagliano, M.~Heindl, P.~Marquetand, M.~F. S.~J. Menger, M.~Oppel,
  F.~Plasser, S.~Polonius, M.~Ruckenbauer, Y.~Shu, D.~G. Truhlar, L.~Zhang,
  P.~Zobel and L.~Gonz{\'a}lez, \emph{{SHARC3.0}: Surface Hopping Including
  Arbitrary Couplings — Program Package for Non-Adiabatic Dynamics},
  https://sharc-md.org/, 2023\relax
\mciteBstWouldAddEndPuncttrue
\mciteSetBstMidEndSepPunct{\mcitedefaultmidpunct}
{\mcitedefaultendpunct}{\mcitedefaultseppunct}\relax
\EndOfBibitem
\bibitem[Mai \emph{et~al.}(2015)Mai, Marquetand, and
  Gonz{\'a}lez]{mai2015general}
S.~Mai, P.~Marquetand and L.~Gonz{\'a}lez, \emph{Int. J. Quantum Chem.}, 2015,
  \textbf{115}, 1215--1231\relax
\mciteBstWouldAddEndPuncttrue
\mciteSetBstMidEndSepPunct{\mcitedefaultmidpunct}
{\mcitedefaultendpunct}{\mcitedefaultseppunct}\relax
\EndOfBibitem
\bibitem[Drautz(2019)]{drautz2019atomic}
R.~Drautz, \emph{Phys. Rev. B}, 2019, \textbf{99}, 014104\relax
\mciteBstWouldAddEndPuncttrue
\mciteSetBstMidEndSepPunct{\mcitedefaultmidpunct}
{\mcitedefaultendpunct}{\mcitedefaultseppunct}\relax
\EndOfBibitem
\bibitem[Batatia \emph{et~al.}(2025)Batatia, Batzner, Kov{\'a}cs, Musaelian,
  Simm, Drautz, Ortner, Kozinsky, and Cs{\'a}nyi]{batatia2025design}
I.~Batatia, S.~Batzner, D.~P. Kov{\'a}cs, A.~Musaelian, G.~N. Simm, R.~Drautz,
  C.~Ortner, B.~Kozinsky and G.~Cs{\'a}nyi, \emph{Nat. Mach. Intell.}, 2025,
  1--12\relax
\mciteBstWouldAddEndPuncttrue
\mciteSetBstMidEndSepPunct{\mcitedefaultmidpunct}
{\mcitedefaultendpunct}{\mcitedefaultseppunct}\relax
\EndOfBibitem
\bibitem[Becke(1988)]{becke1988density}
A.~D. Becke, \emph{Phys. Rev. A}, 1988, \textbf{38}, 3098\relax
\mciteBstWouldAddEndPuncttrue
\mciteSetBstMidEndSepPunct{\mcitedefaultmidpunct}
{\mcitedefaultendpunct}{\mcitedefaultseppunct}\relax
\EndOfBibitem
\bibitem[Becke(1993)]{becke1993new}
A.~D. Becke, \emph{J. Chem. Phys.}, 1993, \textbf{98}, 1372--1377\relax
\mciteBstWouldAddEndPuncttrue
\mciteSetBstMidEndSepPunct{\mcitedefaultmidpunct}
{\mcitedefaultendpunct}{\mcitedefaultseppunct}\relax
\EndOfBibitem
\bibitem[Grimme \emph{et~al.}(2011)Grimme, Ehrlich, and
  Goerigk]{grimme2011effect}
S.~Grimme, S.~Ehrlich and L.~Goerigk, \emph{J. Comput. Chem.}, 2011,
  \textbf{32}, 1456--1465\relax
\mciteBstWouldAddEndPuncttrue
\mciteSetBstMidEndSepPunct{\mcitedefaultmidpunct}
{\mcitedefaultendpunct}{\mcitedefaultseppunct}\relax
\EndOfBibitem
\bibitem[Grimme(2006)]{grimme2006semiempirical}
S.~Grimme, \emph{J. Chem. Phys.}, 2006, \textbf{124}, 034108\relax
\mciteBstWouldAddEndPuncttrue
\mciteSetBstMidEndSepPunct{\mcitedefaultmidpunct}
{\mcitedefaultendpunct}{\mcitedefaultseppunct}\relax
\EndOfBibitem
\bibitem[Dunlap \emph{et~al.}(1979)Dunlap, Connolly, and
  Sabin]{dunlap1979first}
B.~Dunlap, J.~Connolly and J.~Sabin, \emph{J. Chem. Phys.}, 1979, \textbf{71},
  4993--4999\relax
\mciteBstWouldAddEndPuncttrue
\mciteSetBstMidEndSepPunct{\mcitedefaultmidpunct}
{\mcitedefaultendpunct}{\mcitedefaultseppunct}\relax
\EndOfBibitem
\bibitem[Berendsen \emph{et~al.}(1987)Berendsen, Grigera, and
  Straatsma]{berendsen1987missing}
H.~J. Berendsen, J.-R. Grigera and T.~P. Straatsma, \emph{J. Phys. Chem.},
  1987, \textbf{91}, 6269--6271\relax
\mciteBstWouldAddEndPuncttrue
\mciteSetBstMidEndSepPunct{\mcitedefaultmidpunct}
{\mcitedefaultendpunct}{\mcitedefaultseppunct}\relax
\EndOfBibitem
\bibitem[Malde \emph{et~al.}(2011)Malde\emph{et~al.}]{malde2011automated}
A.~K. Malde \emph{et~al.}, \emph{J. Chem. Theory Comput.}, 2011, \textbf{7},
  4026--4037\relax
\mciteBstWouldAddEndPuncttrue
\mciteSetBstMidEndSepPunct{\mcitedefaultmidpunct}
{\mcitedefaultendpunct}{\mcitedefaultseppunct}\relax
\EndOfBibitem
\bibitem[Evans and Holian(1985)]{evans1985nose}
D.~J. Evans and B.~L. Holian, \emph{J. Chem. Phys.}, 1985, \textbf{83},
  4069--4074\relax
\mciteBstWouldAddEndPuncttrue
\mciteSetBstMidEndSepPunct{\mcitedefaultmidpunct}
{\mcitedefaultendpunct}{\mcitedefaultseppunct}\relax
\EndOfBibitem
\bibitem[Mark and Nilsson(2001)]{mark2001structure}
P.~Mark and L.~Nilsson, \emph{J. Phys. Chem. A}, 2001, \textbf{105},
  9954--9960\relax
\mciteBstWouldAddEndPuncttrue
\mciteSetBstMidEndSepPunct{\mcitedefaultmidpunct}
{\mcitedefaultendpunct}{\mcitedefaultseppunct}\relax
\EndOfBibitem
\bibitem[Larsen \emph{et~al.}(2017)Larsen, Mortensen, Blomqvist, Castelli,
  Christensen, Dułak, Friis, Groves, Hammer,
  Hargus,\emph{et~al.}]{larsen2017atomic}
A.~H. Larsen, J.~J. Mortensen, J.~Blomqvist, I.~E. Castelli, R.~Christensen,
  M.~Dułak, J.~Friis, M.~N. Groves, B.~Hammer, C.~Hargus \emph{et~al.},
  \emph{J. Phys. Condens. Matter}, 2017, \textbf{29}, 273002\relax
\mciteBstWouldAddEndPuncttrue
\mciteSetBstMidEndSepPunct{\mcitedefaultmidpunct}
{\mcitedefaultendpunct}{\mcitedefaultseppunct}\relax
\EndOfBibitem
\bibitem[Heydari \emph{et~al.}(2023)Heydari, Raniolo, Livi, and
  Limongelli]{heydari2023transferring}
S.~Heydari, S.~Raniolo, L.~Livi and V.~Limongelli, \emph{Commun. Chem.}, 2023,
  \textbf{6}, 13\relax
\mciteBstWouldAddEndPuncttrue
\mciteSetBstMidEndSepPunct{\mcitedefaultmidpunct}
{\mcitedefaultendpunct}{\mcitedefaultseppunct}\relax
\EndOfBibitem
\bibitem[Shu \emph{et~al.}(2022)Shu, Zhang, Chen, Sun, Huang, and
  Truhlar]{Shu2022}
Y.~Shu, L.~Zhang, X.~Chen, S.~Sun, Y.~Huang and D.~G. Truhlar, \emph{J. Chem.
  Theory Comput.}, 2022, \textbf{18}, 1320–1328\relax
\mciteBstWouldAddEndPuncttrue
\mciteSetBstMidEndSepPunct{\mcitedefaultmidpunct}
{\mcitedefaultendpunct}{\mcitedefaultseppunct}\relax
\EndOfBibitem
\end{mcitethebibliography}
\end{document}